\documentclass[aps,prl,10pt,twocolumn,superscriptaddress,longbibliography,nofootinbib]{revtex4-2}
\usepackage{times}
\usepackage{graphicx}
\usepackage{amsmath,amssymb}
\usepackage{color}
\usepackage[colorlinks=true,linkcolor=blue,citecolor=blue,urlcolor=blue]{hyperref}

\begin{document}

\title{Bosonic Laughlin and Moore-Read states from non-Chern flat bands}
\author{Hongyu Lu}
\affiliation{New Cornerstone Science Lab, Department of Physics, The University of Hong Kong, Hong Kong, China}
\affiliation{HK Institute of Quantum Science \& Technology, The University of Hong Kong, Hong Kong, China}
\affiliation{State Key Laboratory of Optical Quantum Materials, The University of Hong Kong, Hong Kong, China}

\author{Wang Yao}
\email{wangyao@hku.hk}
\affiliation{New Cornerstone Science Lab, Department of Physics, The University of Hong Kong, Hong Kong, China}
\affiliation{HK Institute of Quantum Science \& Technology, The University of Hong Kong, Hong Kong, China}
\affiliation{State Key Laboratory of Optical Quantum Materials, The University of Hong Kong, Hong Kong, China}

\begin{abstract}
The rapid advances in the study of fractional Chern insulators (FCIs) raise a fundamental question: is single-particle band topology a prerequisite for many-body topological order at fractional fillings? This question is pertinent given that, although FCIs were initially discovered in flat Chern bands motivated by their topological equivalence to Landau levels, no-go theorems forbid exact flatness, band topology, and gap isolation from coexisting in finite-range hopping lattices.
Here, we numerically demonstrate bosonic Laughlin and Moore-Read FCIs in two types of non-Chern flat bands on honeycomb lattices, for both hard-core and soft-core bosons.
The Moore-Read state constitutes the first example of a non-Abelian FCI in non-Chern band systems.
Moreover, a common adaptiveness of the boson occupations to the band quantum geometry is observed in these FCIs. The purely on-site interactions considered preclude interaction-induced band topology and dispersion, providing a clean scenario where quantum geometry completely replaces kinetic energy in the interplay with many-body interactions. Our work significantly expands the potential platforms for (non-)Abelian FCIs and reveals new aspects of many-body correlations in the quantum geometry landscape.
\end{abstract}

\date{\today}
\maketitle

\noindent{\textcolor{blue}{\it Introduction.}---}
The recent experimental discovery of fractional Chern insulators (FCI) in correlated moir\'e materials ~\cite{spanton2018observation,Xie2021_topological_CDW_TBG,Cai2023_signature_fqah_mote2,Park2023_observation_fqah_mote2, Zeng2023_thermo_evidence_fqah_mote2,Xu2023_Observation_FQAH_tMote2,Lu2024_FQAH_multilayer_graphene},
together with quantum simulations of their bosonic analogues in artificial lattices~\cite{Julian2023photonFQH, Wang2024_FQH_photon}, have sparked widespread interest in this exotic topological matter.
The FCI phase was first conceived theoretically by translating the integer-to-fractional hierarchy of the quantum Hall effect in Landau levels (LLs) to lattice systems~\cite{Tang2011_FCI, Sun2011_FCI, Sheng2011_FCI, Neupert2011_FCI, Regnault2011_FCI,Xiao2011_quantum_hall}.
Initial explorations naturally focused on the scenario of isolated flat Chern band that resembles the LLs in topology
and energy landscape~\cite{Tang2011_FCI, Sun2011_FCI, Sheng2011_FCI, Neupert2011_FCI, Regnault2011_FCI,Xiao2011_quantum_hall}.
With kinetic energy quenched, quantum geometry becomes the dominant single particle landscape for interplay with many-body interaction.
Having ideal (LL-like) quantum geometry is favorable for stabilizing FCI states~\cite{Parameswaran2012_FCI_algebra,Roy2014_geometry,Classen2015_position_momentum_duality,Ledwith2020_FCI_tbg,Mera2021_geometry_topology,Wang2021_exact_landau_level_description,Ledwith2023_Vortexability}, although not a strict requirement for their existence in a flat Chern band~\cite{Bauer2022_FCI_nonLL,Shavit2024_fci_farfrom_ideal,Grossi2025_gradient_based_search_FCI}.
FCIs have also been found in models where interaction strength well exceeds the band gap isolating the flat Chern band~\cite{Sheng2011_FCI,Kourtis2014_FCI_strong_interactions}.
Notably, in some cases, the Chern bands themselves are not present in the non-interacting limit but instead arise from interaction-driven band renormalization~\cite{Simon2015_FCI_zero_curvature,Kourtis2018_FCI_CDW_trivial_band, Dong2024_FQAH_rhombohedral,Zhou2024_FQAH_rhomohedral,Dong2024_AHC_rhombohedral,Yu2024_multiband_fci}.

The study of FCIs has also moved beyond the flat-Chern-band paradigm along different lines~\cite{Hu2018_fractional_excitonic_insulator,Gassner2025_FEI,Yang2025_FQAH_SFB,Lin2025_FCI_isolated_trivial_band,yang2025_fci_divergent_quantum_geometry}, partly motivated by no-go theorems stating that exact flatness, band topology, and gap isolation cannot be simultaneously satisfied in a finite-range hopping lattice~\cite{Chen_2014_impossibility_exact_flat_chernband,Dubail2015_nogo}.
Latest findings of fermionic Laughlin states in non-Chern flat band models~\cite{Yang2025_FQAH_SFB,Lin2025_FCI_isolated_trivial_band,yang2025_fci_divergent_quantum_geometry,Liu2025topologicalorderbandtopology}  have markedly expanded the horizon.
In particular, exact band flatness, realized via frustrated hopping in bipartite lattices atop trivial or ill-defined band topology~\cite{Yang2025_FQAH_SFB,Lin2025_FCI_isolated_trivial_band,yang2025_fci_divergent_quantum_geometry,Bergman2008_band_touching,Rhim2019_SFB}, leaves quantum geometry as the sole single-particle landscape to consider.
This is a context where our understanding of many-body effects can be reshaped, as exemplified in flat-band superfluidity~\cite{torma_review2022}.
Indeed, the emergence of the Laughlin state in a trivial flat band is rooted in inhomogeneous momentum-space fermion occupation correlated with the quantum metric landscape, enabling a finite occupation-weighted Berry flux despite the vanishing band Chern number~\cite{Lin2025_FCI_isolated_trivial_band,Liu2025topologicalorderbandtopology}.
On the other hand, interactions for such spinless fermions can in principle renormalize the quantum geometry, dispersion, and even band topology.

Bosonic systems with onsite interaction only in the non-Chern flat band lattices provide a cleaner stage, free from kinetic dispersion or interaction-driven band renormalization, for ultimately unraveling the fundamental role of quantum geometry in many-body correlations. We aim to explore here whether the quantum geometry landscape can serve as a general organizing principle for the emergence of FCI in non-Chern bands, encompassing both the Abelian ones and, more ambitiously, non-Abelian FCIs~\cite{MR_nonabelian1991, Read_Rezayi1999,Wu2012_zoology_FCI, Wang2012_nonabelian, Kapit2013_fci_multibody, Liu2013_nonabelian_FCI, Raul_exciton_FCI}.
A central question is whether non-Abelian FCIs,
which are of paramount importance both fundamentally and for fault-tolerant quantum computation~\cite{Nayak_topological_quantum_computation}, can be realized in the absence of band topology, given that they are always more fragile than Abelian ones even in Chern bands and LLs.
Addressing these issues is critical for uncovering the genuinely essential ingredients for FCI emergence, broadening the potential experimental platforms, and revealing new forms of many-body correlations in the quantum geometry landscape.

\begin{figure*}[htp!]
	\centering
	\includegraphics[width=\textwidth]{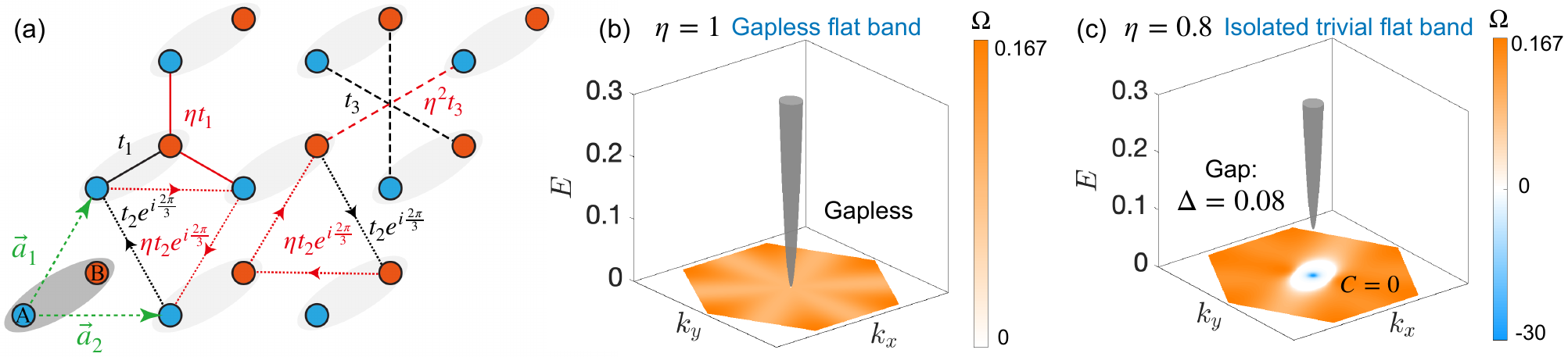}
	\caption{\textbf{Non-interacting model.} (a) Two orbital model on honeycomb lattice including up to third NN hoppings with $-t_1=t_2=t_3=1$.
    (b) When $\eta=1$, the lowest band has exact flatness and a
    quadratic touching point to a dispersive band (grey) at $\Gamma$. (c) When $\eta<1$, the flat band is isolated from the dispersive band by a band gap $\Delta$, while retaining the exact flatness, which necessarily means the band Chern number $C=0$.
    Both (b) the gapless flat band and (c) the isolated trivial flat band are color-coded by their Berry curvature distribution $\Omega(\bf k)$.
    }
	\label{fig_model}
\end{figure*}

Using ED and iDMRG calculations, we demonstrate both Abelian and non-Abelian bosonic FCIs in two types of non-Chern (exactly) flat bands on a honeycomb lattice: a gapless flat band, and a gapped flat band with trivial topology.
At $\nu=1/2$ filling of the
gapless flat band, we identify the bosonic Laughlin state stabilized by onsite interactions from hard-core limit down to arbitrarily small strength, with the weak-interaction stability enabled by the exact flatness of the band.
Under lattice parameters that realize a gapped trivial flat band, we also find the Laughlin FCI across the entire interaction range, from soft-core bosons in the isolated band limit (onsite interaction $\ll$ band gap) to hard-core bosons.
Furthermore, we discover bosonic Moore-Read states in both kinds of non-Chern flat bands at $\nu=1$ filling, the first example of non-Abelian FCI in non-Chern band systems.
These (non-)Abelian FCIs stabilized by purely onsite interactions highlight their genuine non-Chern-band origin.
Remarkably, we observe a common quantum-geometry adaptiveness in both Laughlin and Moore–Read states, albeit with distinct manifestations in the gapped trivial band and in the gapless case: in regions where the flat band is well isolated, the boson occupation correlates with the flat-band quantum metric; in the vicinity of a band degeneracy, the two-band quantum metric takes over.

\noindent{\textcolor{blue}{\it Models and methods.}---}
The two-orbital model on the honeycomb lattice is:
\begin{equation}
\begin{aligned}
    H={}&\sum_{ij}t_{ij}(b_i^\dagger b_j^{\ }+h.c.)+\mu\sum_i n_i\\
       &+\sum_n\frac{U_n}{n!}\sum_i(b_i^\dagger)^n(b_i^{\ })^n.
\end{aligned}
\end{equation}
The non-interacting part including NN, second NN, and third NN hoppings is illustrated in Fig.\ref{fig_model} (a). The hopping amplitude $\left| t_{ij}\right|=1$ serves as the energy unit. When $\eta=1$, this non-interacting part describes a gapless flat band model with the lower band being exactly flat [Fig.\ref{fig_model}(b)] (the zero modes are shown in the \hyperref[sec:zero-modes]{supplementary information (SI)}),
and a band touching to a dispersive band dictated by real-space topology~\cite{Bergman2008_band_touching,Rhim2019_SFB,Rhim2021_SFB}.
The Bloch wavefunction is non-analytic under any gauge choice
at the touching point at $\Gamma$,
while the quantum geometry elsewhere is relatively uniform,
and we show the $\Omega$ distribution of the gapless flat band in Fig.\ref{fig_model} (b).
When $\eta<1$, the flat band is isolated from the upper band by a band gap $\Delta\approx2(\eta-1)^2$ [Fig.\ref{fig_model} (c)], while retaining the exact flatness~\cite{Lin2025_FCI_isolated_trivial_band}. This necessarily means the band Chern number $C=0$.
The $\Omega$ distribution of the $C=0$ flat band has a negative peak at $\Gamma$, and a positive and relatively uniform background in the rest of the BZ [Fig.\ref{fig_model} (c)]. $\mu=(2+\eta^2)t$ fixes the non-interacting flat band at zero energy.
In the main text, we will focus on the effect of onsite interactions by considering both the hard-core (with some $U_n$ being infinite) and soft-core (finite $U_n$) cases, and will discuss the NN interactions in the \hyperref[figs_Laughlin_sfb_V1]{SI}.

\begin{figure}[t]
	\centering
	\includegraphics[width=\columnwidth]{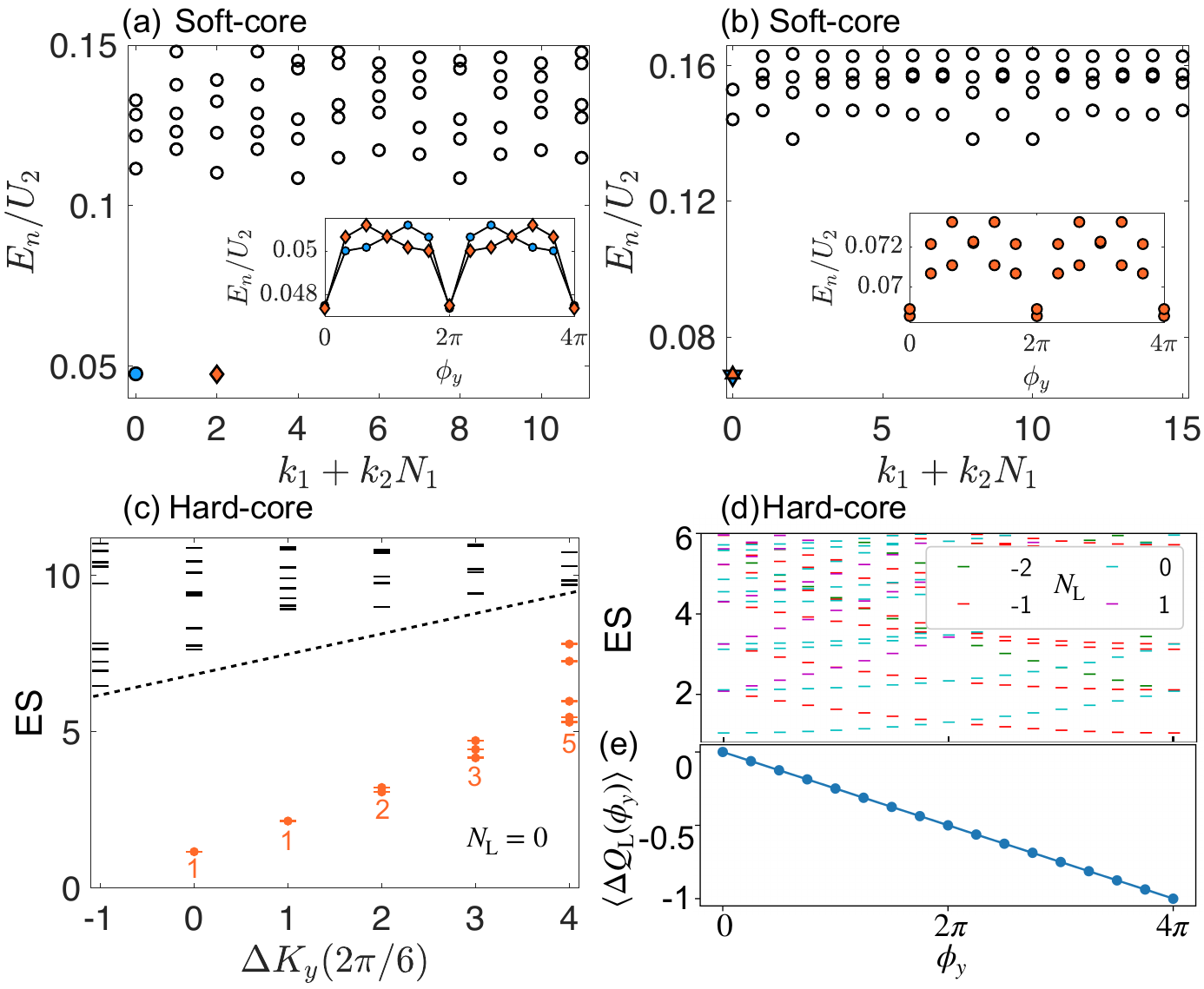}
	\caption{\textbf{Laughlin FCI in the gapless flat band with arbitrarily small and large onsite interactions.} We consider $\nu=1/2$ filling of the gapless flat band ($\eta=1$).
    ED spectra of soft-core bosons with a very small $U_2=0.001$ interaction from $N=4\times3\times2$ and $N=4\times4\times2$ tori are shown in panels (a,b), respectively.
    For each system size, the spectral flow of the well gapped two-fold ground states is shown in the inset.
    The iDMRG results of hard-core bosons ($U_{n\geq2}=\infty$) are shown in panel (c-e).
    In panel (c), the momentum-resolved entanglement spectrum (ES) of the charge sector ($N_\mathrm{L}=0$) with the largest eigenvalue of the reduced density matrix shows the characteristic counting $\{1,1,2,3,5,...\}$ of the edge
conformal field theory. We further show (d) the evolution of ES of different charge sectors ($N_\mathrm{L}=-2,-1,0,1$) and (e) charge pumping results by adiabatically inserting $4\pi$ fluxes.
    }
	\label{fig_laughlin_sfb}
\end{figure}

We label the number of two-orbital unit cells along the two primitive vectors as $N_1$ ($\vec{a}_1$) and $N_2$ ($\vec{a}_2$) with a total of $N_1\times N_2\times2$ sites.
The two-band ED simulations only take the soft-core case such that the Hilbert space with $N_b$ bosons is $\mathrm{C}(N_1\times N_2\times2+N_b-1,N_b)$.
The iDMRG simulations are based on the real-space basis with the maximum onsite occupation as $n_\mathrm{max}$ and the local Hilbert-space dimension is $d=n_\mathrm{max}+1$.
The hard-core condition here is realized by limiting $n_\mathrm{max}$, which means $U_{n>n_\mathrm{max}}=\infty$. To address the soft-core case in iDMRG, we do not need infinite local Hilbert space since the maximal number of bosons in the iDMRG unit cell is $N_1\times N_2\times\nu$, bounded by filling factor $\nu$ of the flat band.
We keep up to $D=3000$ bond dimensions in iDMRG simulations with maximum truncation errors around $10^{-5}$ and $10^{-6}$ for the Abelian and non-Abelian FCIs respectively.

\noindent{\textcolor{blue}{\it Bosonic Laughlin FCI in the gapless flat band.}---}
We first consider $\nu=1/2$ filling of the gapless flat band ($\eta=1$) and start with soft-core bosons and only very small $U_2=0.001$ ($U_{n>2}=0$) interaction.
The spectra from two-orbital ED simulations of $N=4\times3\times2$ and $N=4\times4\times2$ tori are shown in Fig.\ref{fig_laughlin_sfb} (a,b), respectively.
The two-fold ground states (with their spectral flow in the insets) are well gapped from the excited states and their momenta are in agreement with the generalized Pauli principle for the bosonic 1/2 FCI on torus~\cite{Haldane1991_GPP, Bernevig2008_jack_polynomials, Regnault2011_FCI}.
The many-body Chern number of the ground states gives the fractional quantized Hall conductivity $\sigma_\mathrm{H}=\frac{1}{2}\frac{e^2}{h}$, further confirming the bosonic Laughlin-like FCI.
In fact, due to the exact flatness of the gapless flat band, arbitrarily small $U_2$ could stabilize the FCI, which has not been numerically reported in Chern-band systems.
The considered $U_2=0.001$ is already very small and the low-energy spectra have well converged with decreasing $U_2$ in the unit of $E_n/U_2$ (c.f. \hyperref[figs_ed_different_U2_sfb]{SI}).

We then go to the opposite limit, i.e. hard-core bosons with $U_{n\geq2}=\infty$. The iDMRG results with circumference $N_1=6$ are shown in Fig.\ref{fig_laughlin_sfb}(c-e).
The momentum-resolved entanglement spectrum (ES) shows the characteristic edge-mode counting $\{1,1,2,3,5,...\}$, in agreement with the edge conformal field theory, supporting the topological nature of the ground state [Fig.\ref{fig_laughlin_sfb}(c)]~\cite{LiHaldane2008_ES, Cincio2013_topological_order_idmrg}.
After adiabatically inserting $4\pi$ fluxes, the ES goes back to itself while shifted by one charge sector and there is one boson pumped from the left to the right side of the system, suggesting the FCI ground state with $\sigma_\mathrm{H}=\frac{1}{2}\frac{e^2}{h}$~\cite{Laughlin1981_hall_conductivity}.
Therefore, we have shown that the Laughlin FCI could be stabilized by arbitrarily small and large onsite two-body interactions. In the \hyperref[figs_Laughlin_sfb_V1]{SI}, we further show that this FCI could survive with very large NN interactions as well.

\begin{figure}[htp!]
	\centering
	\includegraphics[width=\columnwidth]{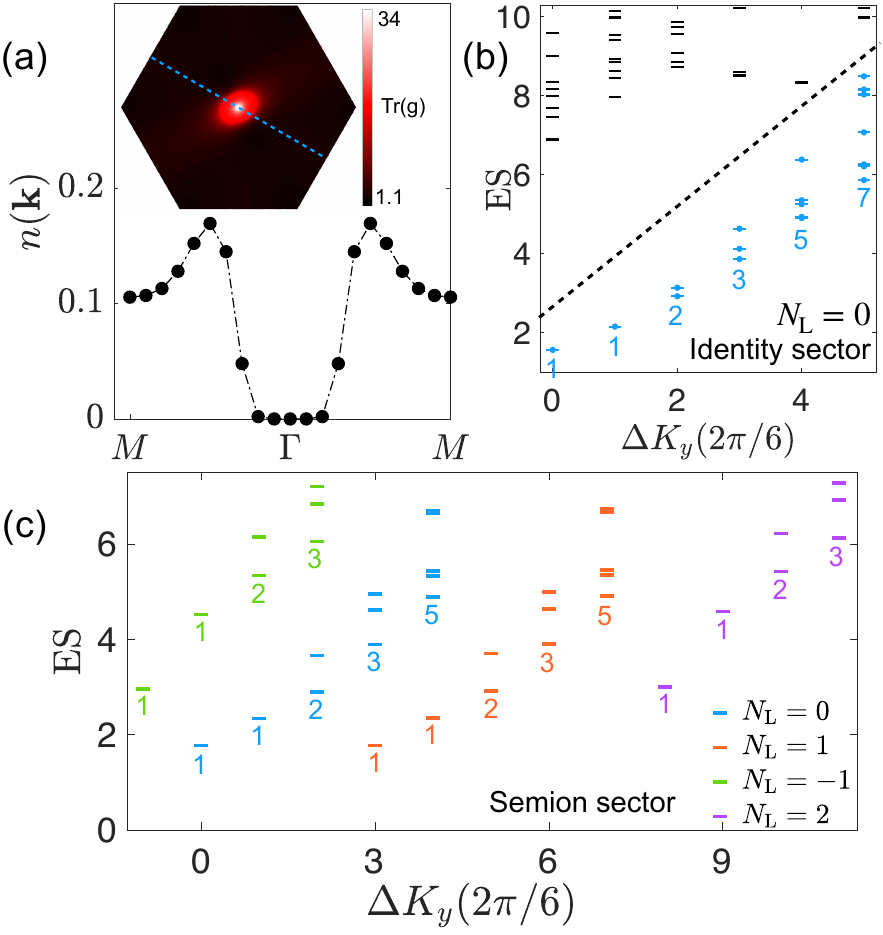}
	\caption{\textbf{Laughlin FCI in the isolated $C=0$ flat band with soft-core bosons.} The iDMRG results at $\nu=1/2$ filling of  the isolated flat band ($\eta=0.8$ with band gap $\Delta=0.08$) with only $U_2=0.02$.
	The inset of panel (a) shows the trace of quantum metric of the trivial flat band with a peak at $\Gamma$, and the blue dashed line shows the path along which we compute the momentum-space boson distribution $n(\mathbf{k})$ in (a).  	$n(\mathbf{k})$ is measured from the ground state in the identity sector for example and the bosons tend to avoid occupying around $\Gamma$.
		(b) The ES of the $N_\mathrm{L}=0$ charge sector of the ground state in the identity sector, showing the characteristic edge-mode counting $\{1,1,2,3,5,7,... \}$.
		(c) The ES of the ground state in the semion sector, where all charge sectors show the same degeneracy pattern and the ES is symmetric about $N_\mathrm{L}=0.5$.
	}
	\label{fig_laughlin_trivial_band}
\end{figure}

\noindent{\textcolor{blue}{\it Bosonic Laughlin FCI in the gapped $C=0$ flat band.}---}
At $\eta=0.8$, the (exactly) flat band is gapped with $\Delta=0.08$.
Like the Berry curvature, the trace of quantum metric also shows a peak at $\Gamma$ [inset of Fig.~\ref{fig_laughlin_trivial_band}(a)].
When spinless fermions form FCI in such $C=0$ band~\cite{Lin2025_FCI_isolated_trivial_band}, the nearest-neighbor repulsion tends to drive carriers away from such peak, thereby occupying BZ region with relatively uniform quantum geometry.
However, such nearest-neighbor repulsion can also renormalize the band and introduce effective dispersion.
The coordination of correlation and non-trivial quantum geometry in exactly flat bands with no band renormalization, and the exitence of FCIs therein, are unclear.
We address this open issue by considering $\nu=1/2$ filling of the isolated trivial flat band, and starts with the onsite two-body repulsion that is much weaker than the band gap.

The soft-core-boson condition is realized with onsite $n_\mathrm{max}=N_1\times N_2\times\nu$ since each iDMRG unit cell can not have more bosons. Without explicit interactions, the system is basically non-interacting and the ground state is topologically trivial.
We will consider $U_2=0.02$ ($U_{n>2}=0$), which is much smaller than the band gap and thus ensuring the isolated-band limit.
The results of $N_1=6$ (circumference of the infinite cylinder) and $N_2=1$ (we have checked that there will be no translation symmetry breaking along $\vec{a}_2$ with only small $U_2$) are shown in Fig. \ref{fig_laughlin_trivial_band}.
With random initial conditions and optimization, we always obtain two nearly degenerate ground states~\cite{Cincio2013_topological_order_idmrg,Zhu2015_MR_DMRG}, which actually refer to the two topological sectors of the 1/2 Laughlin state.

To reveal the mechanism of forming this FCI in an isolated trivial band, we compute the momentum-space boson distribution $n(\mathbf{k})=\sum_\alpha\sum_j\langle b_{i,\alpha}^\dagger b_{j,\alpha}^{\ } \rangle e^{-i\mathbf{k}\mathbf{r}_{ij}}$ by considering a finite region of the infinite cylinder, where $\alpha$ refers to the A/B sublattice.
The results from both topological sectors show that $n(\mathbf{k})$ is strongly suppressed when approaching $\Gamma$, suggesting that this FCI is realized by occupying the BZ away from $\Gamma$ with relatively uniform $\Omega$.
The $n(\mathbf{k})$ from the identity sector along one k-path is shown in Fig.\ref{fig_laughlin_trivial_band} (a).
This strongly supports that the correlation could coordinate solely with the non-trivial quantum geometry in exact flat bands without any single-particle dispersion.
The ES of the charge sector with largest eigenvalue of the reduced density matrix from the identity-sector ground state is further shown in Fig.\ref{fig_laughlin_trivial_band} (b), with the characteristic edge mode counting $\{1,1,2,3,5,7,... \}$.
Moreover, we show the ES from the semion-sector ground state in Fig.\ref{fig_laughlin_trivial_band} (c)
with all the charge sectors having the same degeneracy pattern $\{1,1,2,3,5,... \}$.
The ES is symmetric about $N_\mathrm{L}=0.5$, confirming the semion anyon at each edge of the cylinder, carrying the fractional charge $e/2$~\cite{Wen1990_chiral_luttinger_liquid,LiHaldane2008_ES,Cincio2013_topological_order_idmrg}.

We have also explored the Laughlin state in the gapped trivial flat band with hard-core bosons, and it is adiabatically connected to that in the gapless flat band (see the \hyperref[figs_Laughlin_eta0d8_hardcore]{SI}).

\begin{figure}[htp!]
	\centering
	\includegraphics[width=\columnwidth]{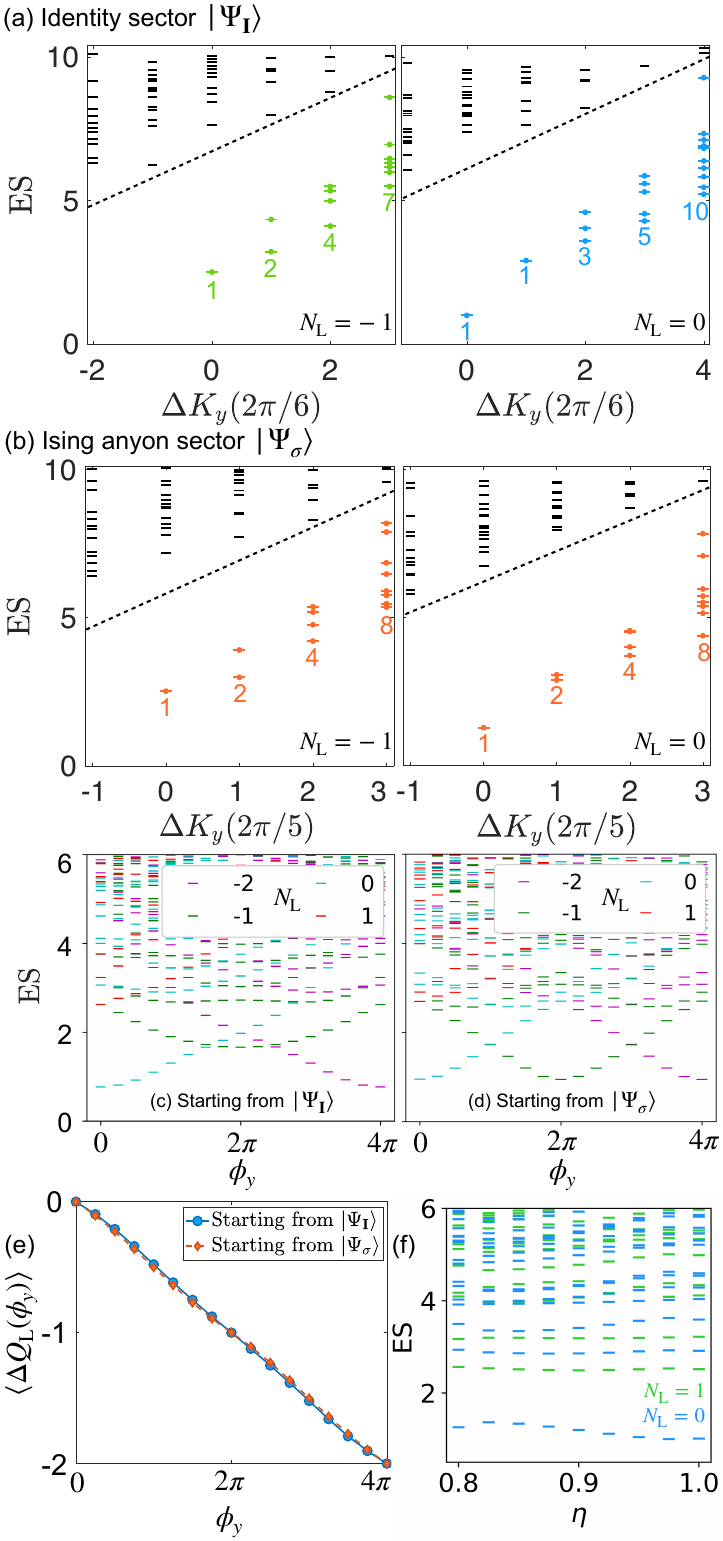}
	\caption{\textbf{Moore-Read state in non-Chern flat band.} We show the iDMRG results at $\nu=1$ of the gapless flat band with hard-core bosons ($n_\mathrm{max}=2$) including:
	(a) the ES of $\Psi_\mathbf{I}$ ground state on the $N_1=6$ cylinder, showing the edge mode counting $\{1,2,4,7,... \}$ and  $\{1,1,3,5,10... \}$ for the $N_\mathrm{L}=-1$ and $N_\mathrm{L}=0$ charge sectors, respectively;
	(b) the ES of the $\Psi_\sigma$ ground state on the $N_1=5$ cylinder, showing the same degeneracy pattern $\{1,2,4,8,... \}$ in all charge sectors;
	the evolution of ES after flux insertion, starting from (c) the $\Psi_\mathbf{I}$  or (d) the $\Psi_\sigma$ ground state.
    (e) charge pumping results starting from the $\Psi_\mathbf{I}$ or the $\Psi_\sigma$ ground state; (f) adiabatic evolution of the ES between the Moore-Read state (identity sector) in the gapless flat band and that in the gapped trivial flat band.
    }
	\label{fig_MR_nonchern}
\end{figure}

\noindent{\textcolor{blue}{\it Bosonic Moore-Read state in non-Chern flat band.}---}
Furthermore, we find the existence of the MR state in the non-Chern flat bands, where we mainly consider the $\nu=1$ filling with three-body hard-core bosons ($n_\mathrm{max}=2$), which is numerically easier than the soft-core case for this high filling.
We begin with the results in gapless flat band.
The ES of the identity sector $\Psi_\mathbf{I}$ ground state is shown in Fig.\ref{fig_MR_nonchern}(a), with the characteristic edge-mode of counting $\{1,2,4,7,... \}$ for odd charge sectors and $\{1,1,3,5,10... \}$ for even charge sectors. The even-odd difference can be understood from the root configuration ``..020202..'' of $\Psi_\mathbf{I}$ depending on how it is cut~\cite{Ardonne_2005_edge,LiHaldane2008_ES,Bernevig2008_jack_polynomials,Papic2011_entanglement_excitation,Liu2012_edge_mode_es,Zhu2015_MR_DMRG}.
To obtain the Ising anyon sector $\Psi_\sigma$, we consider an iDMRG unit cell with $N_1=5$ and $N_2=1$ such that only the root configuration ``..11111..'' is allowed.
The ES of the $\Psi_\sigma$ ground state is shown in Fig.\ref{fig_MR_nonchern}(b), and the edge mode degeneracy pattern is $\{1,2,4,8,... \}$ for all charge sectors, consistent with the analytical prediction of Ising anyon primary field~\cite{Ardonne_2005_edge,Liu2012_edge_mode_es,Zhu2015_MR_DMRG}.

For the adiabatic flux insertion, the charge pumping results starting from the $\Psi_\mathbf{I}$ or the $\Psi_\sigma$ ground states are shown in Fig.\ref{fig_MR_nonchern}(e).
After inserting $2\pi$ ($4\pi$), there is 1 (2) net charges pumped from the left to the right edge of the systems for both cases. However, the underlying fusion rules of the quasiparticles are different~\cite{Zhu2015_MR_DMRG}.
Starting from the $\Psi_\mathbf{I}$ ground state, the evolution of ES with flux insertion is shown in Fig.\ref{fig_MR_nonchern}(c).
Although there is a net charge pumped after insertion of $\phi_y=2\pi$ flux, the ES does not go back to itself and is different from that at $\phi_y=0$.
This is because the ground state adiabatically evolves into the fermion anyon sector $\Psi_f$, where the ES would exhibit edge-mode counting $\{1,2,4,7,... \}$ for even charge sectors and $\{1,1,3,5,10... \}$ for odd charge sectors. The even-odd counting is shifted by one charge for the $\Psi_\mathbf{I}$ and $\Psi_f$ ground states, which is also manifested from the comparison of different ES counting in the $N_\mathrm{L}=-1,0$ sectors in Fig.\ref{fig_MR_nonchern}(a).
After inserting $4\pi$ flux, the ES goes back to itself with the charge sectors shifted by $2$, and the ground state goes back to the $\Psi_\mathbf{I}$ again.
These relate to the fusion rules: $\mathbf{I}\times f=f$ and $f\times f=\mathbf{I}$.
If starting from the $\Psi_\sigma$, just as the same edge mode counting in different charge sectors (Fig.\ref{fig_MR_nonchern} (b)), the ES would be the same with each integer number of $2\pi$ flux (Fig.\ref{fig_MR_nonchern} (d)). This relates to $\sigma\times f=\sigma$ and the $\Psi_\sigma$ ground state always evolves into itself with flux insertion of $2\pi$, unlike the $\Psi_\mathbf{I}$ and $\Psi_f$ ground states that evolve into each other. This also demonstrates that the Ising anyon  does not
respond to the $U(1)$ charge flux.

At $\nu=1$ filling of the gapped trivial flat band, we have also discovered the Moore-Read state. The characteristic edge ES from iDMRG simulations with hard-core bosons, and the 3-fold quasi-degenerate ground states from one-band ED simulations with soft-core bosons by projecting the onsite three-body interaction, are shown in the \hyperref[sec:moore-read]{SI}.
Interestingly, as shown in Fig.~\ref{fig_MR_nonchern} (f), the ES evolves smoothly between the Moore-Read state in the gapless flat band and that in the gapped trivial flat band, indicating that the Moore-Read states in the two types of non-Chern flat bands are adiabatically connected.

\begin{figure}[htp!]
	\centering
	\includegraphics[width=\columnwidth]{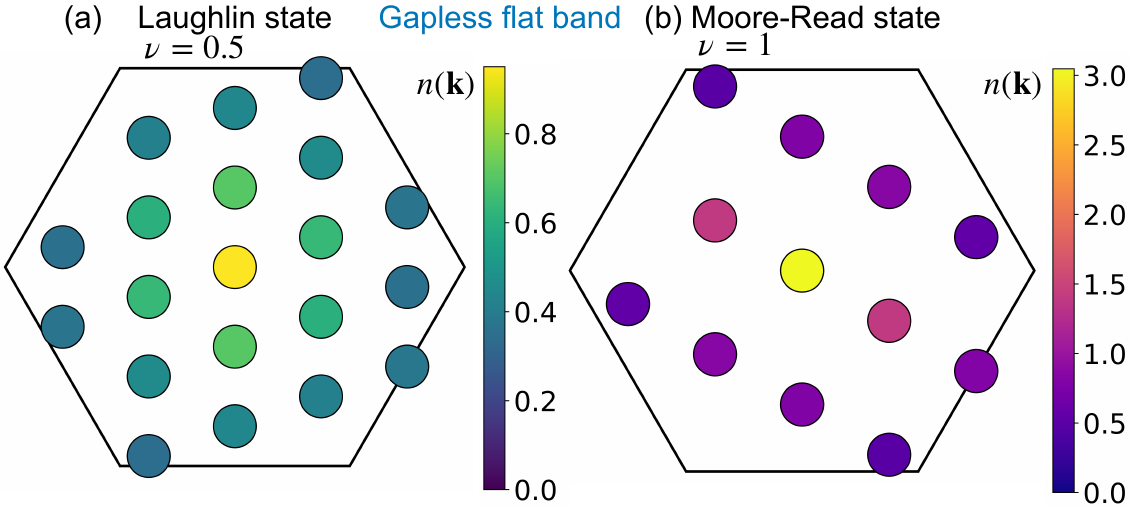}
	\caption{\textbf{Momentum-space boson occupations $n(\mathbf{k})$ in gapless flat band.} (a) Laughlin state at $\nu=0.5$ with $N_b=10$ bosons, and (b) Moore-Read state at $\nu=1$ with $N_b=12$ bosons,  from full ED simulations with hard-core bosons.
    }
	\label{fig_nk_adaptiveness}
\end{figure}

\noindent{\textcolor{blue}{\it Quantum-geometry adaptiveness.}---}
In the gapped trivial band, the depletion of boson occupation at the quantum metric peak at $\Gamma$ is not only observed for the Laughlin state (Fig.~\ref{fig_laughlin_trivial_band}(a)), but also for the Moore-Read state (Fig.~\ref{figs_nk_MR_eta0d8_idmrg}). This could be understood from the fact that the trace of quantum metric characterizes the real-space spread of Bloch particles~\cite{Marzari1997_maximally_localized}, whereas particles prefer smaller overlaps with each other due to the repulsion.
In the gapless case, however, the band touching invalidates
the projected band description in its vicinity and raises the question of whether multi-band quantum geometry shall be considered instead.
To address this question, we calculate the $n(\mathbf{k})$ in the gapless flat band from full ED simulations with hard-core bosons. Interestingly, $n(\mathbf{k})$ for both the Laughlin and Moore–Read FCIs here exhibits opposite features to its gapped counterparts, with prominent peaks at $\Gamma$. This apparent contrast is resolved by noting the vanishing two-band quantum metric, which takes over as the relevant quantity due to band degeneracy in the vicinity of band touching.
More ED results and iDMRG results in the \hyperref[sec:geometry-adaptiveness]{SI} consistently support this common quantum geometry adaptiveness.

\noindent{\textcolor{blue}{\it Discussions.}---}
Our work demonstrates that the bosonic (non-)Abelian FCIs could be genuinely realized in {\it non-Chern} flat band systems with purely onsite interactions (that do not renormalize the exactly flat at the single-particle level), which broadens the understanding of the origins and mechanisms of FCI.  The stability of the FCIs is supported by consistent diagnostics over the accessible system sizes (see the \hyperref[sec:robustness]{SI}), and future efforts of controlled extrapolation to the thermodynamic limit would be beneficial possibly with more advanced numerical methods.
Whereas many-body interactions are often crucial, we provide a distinct perspective centered on flat-band quantum geometry, pointing to FCI physics beyond band topology.
The common quantum-geometry adaptiveness observed in the distinct cases here could have broader implications for the
general exploration of many-body correlations in flat-band systems.

Due to the exact flatness of these non-Chern bands, FCIs could be induced by arbitrarily small onsite interactions~\footnote{Such FCIs in non-Chern flat bands are also robust against finite dispersions and the bandwidth tolerance is comparable to FCIs in the optimized Chern-band Haldane model (see the \hyperref[figs_bandwidth_eta]{SI}).}.
This feature surpasses the limitations of Chern bands, which must have a finite bandwidth in a finite range hopping lattice~\cite{Chen_2014_impossibility_exact_flat_chernband}, thereby requiring a minimum interaction strength to overcome their dispersion.
This enables investigation in a broad range of platforms - including photonic, phononic, excitonic and polaritonic systems~\cite{Wu2016_circuit_qed,
Amo2016_excitonpolariton_lattice,Wang2020_floquet_photoncavity_lattice, Yang2020_photonic_floquet, Clark2020_laughlin_light, Mathew2020_gaugefield_phonon, He2021_flat_band_superradiance, Xia2024_phonon_flat_band,Xie2024_flatband_excitons,Raul_exciton_FCI} - thereby significantly expanding the landscape for exploring bosonic FCIs.
We note that the three-body onsite interaction to stabilize the Moore-Read state can be realized  in optical lattices with ultracold atoms~\cite{Daley2009_atomic_3body_interaction, Julian2023photonFQH}, circuit QED systems~\cite{Hafezi2014_3body_circuit_qed, Wang2024_FQH_photon}, and microwave-dressed polar molecules~\cite{Buchler2007_3body_cold_polar}.
It is also interesting to further investigate whether the non-Abelian phase could be realized in such non-Chern flat bands with purely two-body interactions (see the \hyperref[sec:two-body]{SI}), as reported in Chern-band systems~\cite{Liu2013_nonabelian_FCI,Palm2021_bosonic_pfaffian,Raul_exciton_FCI}.

In an extended Bose-Hubbard model on the Kagome lattice with infinite on-site interaction and necessary competing different-range interactions, the bosonic Laughlin state has been reported~\cite{Zhu2016_FQH_kagome_onethird}, which is a chiral spin liquid in spin language as the model can be mapped onto the spin-1/2 XXZ model~\cite{Kalmeyer1987_CSL,Kumar2014_CSL_XXZ_kagome}.
Upon spontaneous time reversal symmetry breaking, the 1/2 Laughlin state emerges at an integer filling of hard-core bosons per unit cell~\cite{Zhu2016_FQH_kagome_onethird}, which is intrinsically different from the Laughlin states in our scenarios at half flat-band fillings.
To the best of our knowledge, our work also demonstrates the first explicit example of FCIs of
bosons with arbitrary interaction strength beyond the conventional Chern-band paradigm.

While the FCIs in the gapped trivial flat band are robust with over a large range of $\eta$ value, they eventually give way to topologically trivial insulators at sufficiently small $\eta$ where the peak of quantum geometry becomes less sharp (see the \hyperref[sec:phase-diagram]{SI}).
This is a novel example of quantum-geometry-driven topological many-body transition without changing band topology, flat-band dispersion, or interaction strength, and might also offer a new and useful setting for future explorations of continuous transitions beyond the conventional Ginzburg-Landau paradigm~\cite{Wen1993_fqh_insulator_transition,Barkeshli2014_fci_sf,Song2023_deconfined_criticality_duality}.

\begin{acknowledgments}
{\it Acknowledgments}\,---\, We thank Jie Wang and Nicolas Regnault for helpful discussions. We thank Zuzhang Lin, Wenqi Yang, and Dawei Zhai for earlier collaborations.
The work is supported by the National Natural Science Foundation of China (No. 12425406), Research Grant Council of Hong Kong (AoE/P-701/20, HKU SRFS21227S05), and New Cornerstone Science Foundation. We thank Beijing PARATERA Tech Co., Ltd. (https://cloud.paratera.com) for providing HPC resources that supported the research results reported in this paper.
TeNPy library is used in the iDMRG simulations~\cite{Johannes_2024_tenpy}.

\end{acknowledgments}

%

\onecolumngrid
\clearpage

\setcounter{figure}{0}
\setcounter{equation}{0}
\renewcommand{\thefigure}{S\arabic{figure}}
\renewcommand{\theequation}{S\arabic{equation}}
\renewcommand{\theHfigure}{supp.\arabic{figure}}
\renewcommand{\theHequation}{supp.\arabic{equation}}

\section*{Supplementary Information}

Here, we show additional information including: the real space topology underlying the gapless flat band in Sec. A; the Moore-Read state in the gapped trivial flat band in Sec. B; the robustness of Laughlin and Moore-Read states in the non-Chern flat bands in Sec. C; the quantum phase diagram and transition by tuning $\eta$ in Sec. D;  the effect of two-body interactions at $\nu=1$ of the non-Chern flat bands in Sec. E; and the quantum-geometry adaptiveness in Sec. F.

\subsection*{A. Zero modes of the gapless flat band}
\label{sec:zero-modes}

\begin{figure}[htp!]
	\centering
	\includegraphics[width=0.7\textwidth]{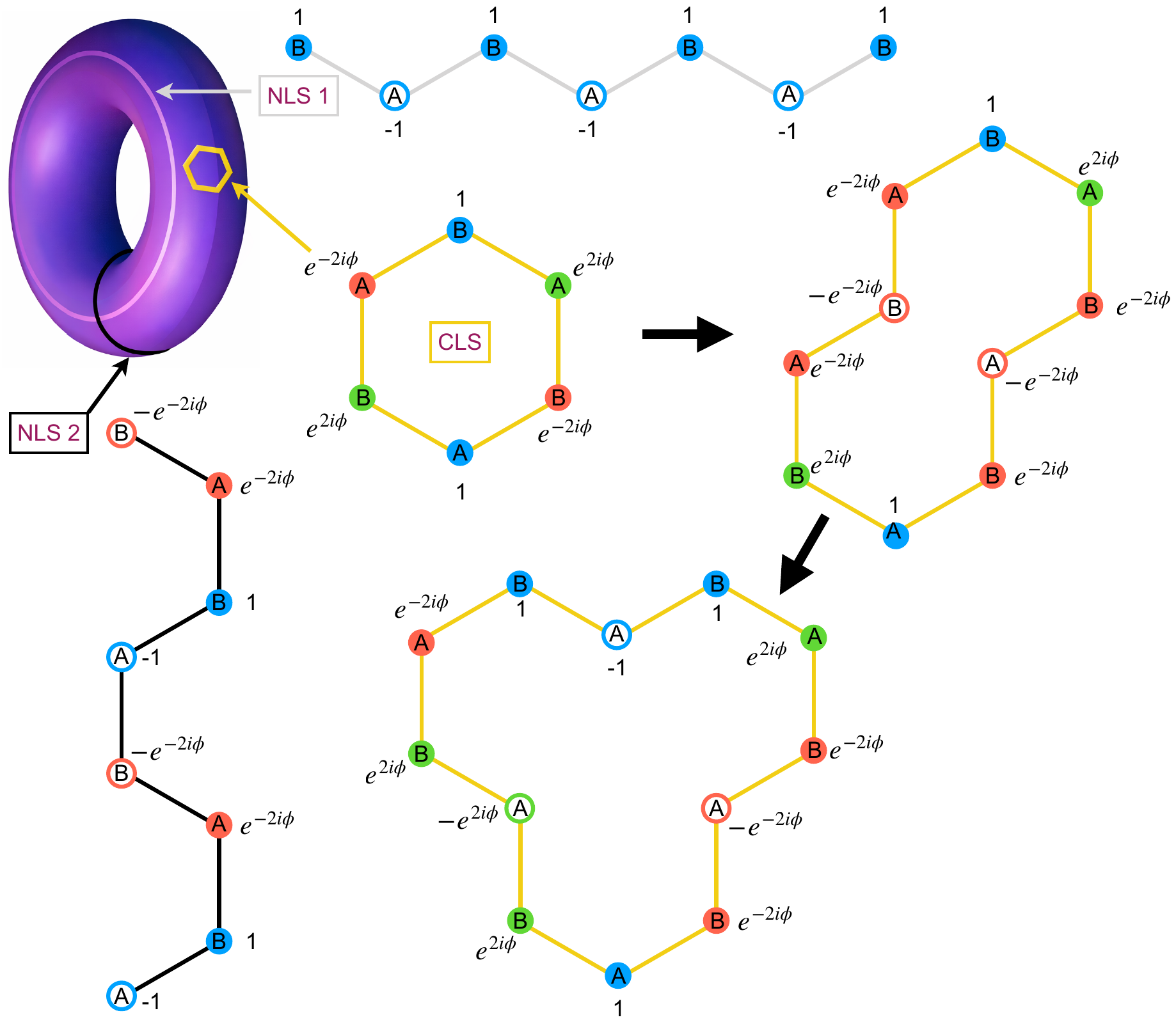}
	\caption{\textbf{Schematics of the compact localized state (CLS) and two non-contractible loop states (NLS) on the torus at $\eta=1$.}
    A and B Circles denote the orbitals with the numbers (color) denoting their probability amplitudes in the CLS or NLS.
    The states from adding two and three adjacent CLS are also shown. There are only $N-1$ linearly independent CLS on torus, and the overall counting of zero modes are consistent with the gapless flat band with a touching point.}
	\label{figs_SFB_CLS_NLS}
\end{figure}

For the gapless flat band, we have mentioned that the Bloch
wavefunction is non-analytic in any gauge choice at the touching point at $\Gamma$.
Here, we further demonstrate the zero modes. As shown in Fig.~\ref{figs_SFB_CLS_NLS}, there are only only $N-1$ linearly independent compact localized states (CLS) on the torus (insufficient to span a flat band), but additionally there are two NLS at zero energy, which are characteristics of the singular flat band~\cite{Rhim2019_SFB}.
All together, there are $N+1$ linearly independent zero modes, dictating an exactly flat band with a single quadratic touching point.
Fig.~\ref{figs_SFB_CLS_NLS} also show the states from adding two and three adjacent CLS.
Clearly, with open boundary condition, adding all CLS centered at all the unit cells leads to the robust boundary mode~\cite{Rhim2019_SFB}.

\subsection*{B. Moore-Read state in the gapped trivial flat band}
\label{sec:moore-read}

\begin{figure}[htp!]
	\centering
	\includegraphics[width=0.8\textwidth]{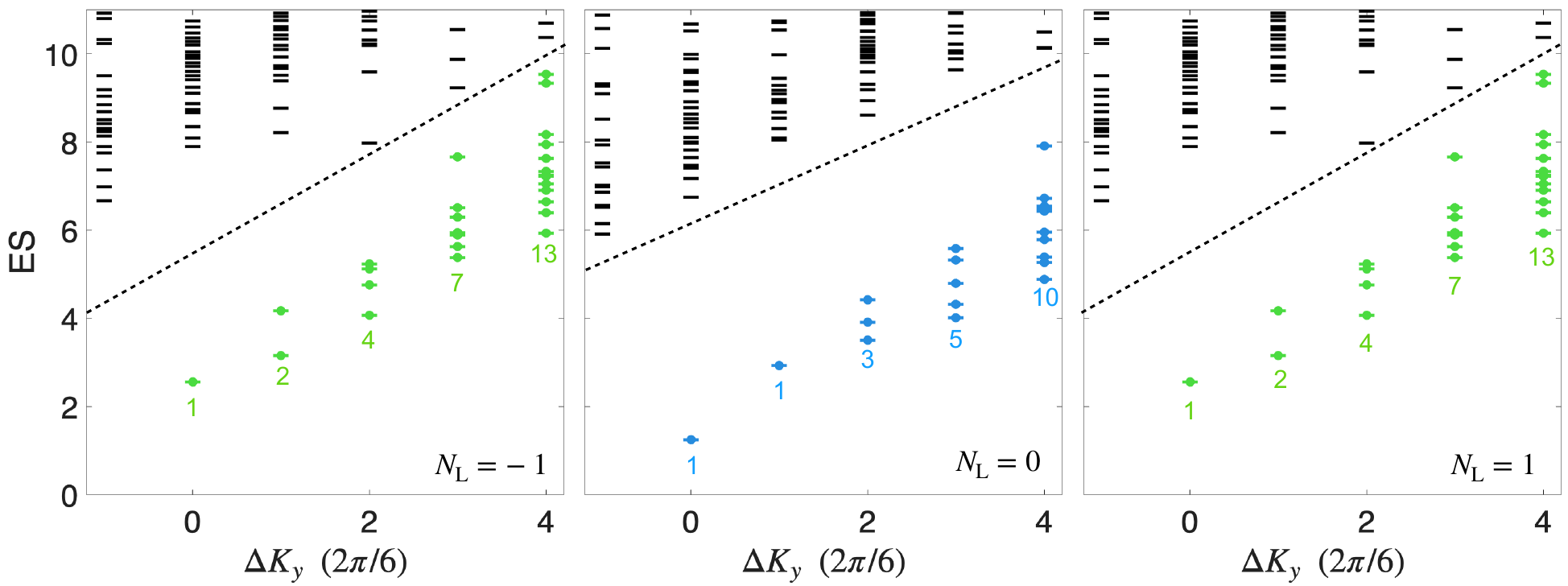}
	\caption{\textbf{Momentum-resolved entanglement spectra of the Moore-Read state in the gapped trivial band.}
    These results are computed in iDMRG simulations at $\nu=1$ filling of the gapped trivial band ($\eta=0.8$) with hard-core bosons ($n_\mathrm{max}=2$) on the circumference $N_y=6$ cylinder. The entanglement spectra are computed from the identity sector of this Moore-Read state. The characteristic edge-mode counting $\{1,2,4,7,13,...\}$ for odd charge sectors and $\{1,1,3,5,10,...\}$ for even charge sectors is in agreement with the edge conformal field theory, supporting the Moore-Read ground state.  }
	\label{figs_ES_MR_hardcore_0d8}
\end{figure}

As mentioned in the main text, the Moore-Read states can be realized in both the gapless flat band and the gapped trivial flat band, and the adiabatic connection of the many-body phase in these two limits is also studied.
As a supplement, we show the momentum-resolved entanglement spectrum of the Moore-Read state in the gapped trivial band from iDMRG simulations on the circumference $N_y=6$ cylinder in Fig.~\ref{figs_ES_MR_hardcore_0d8}.
The characteristic edge-mode counting provides clear evidence for the non-Abelian Moore-Read state.

\begin{figure}[htp!]
	\centering
	\includegraphics[width=0.5\textwidth]{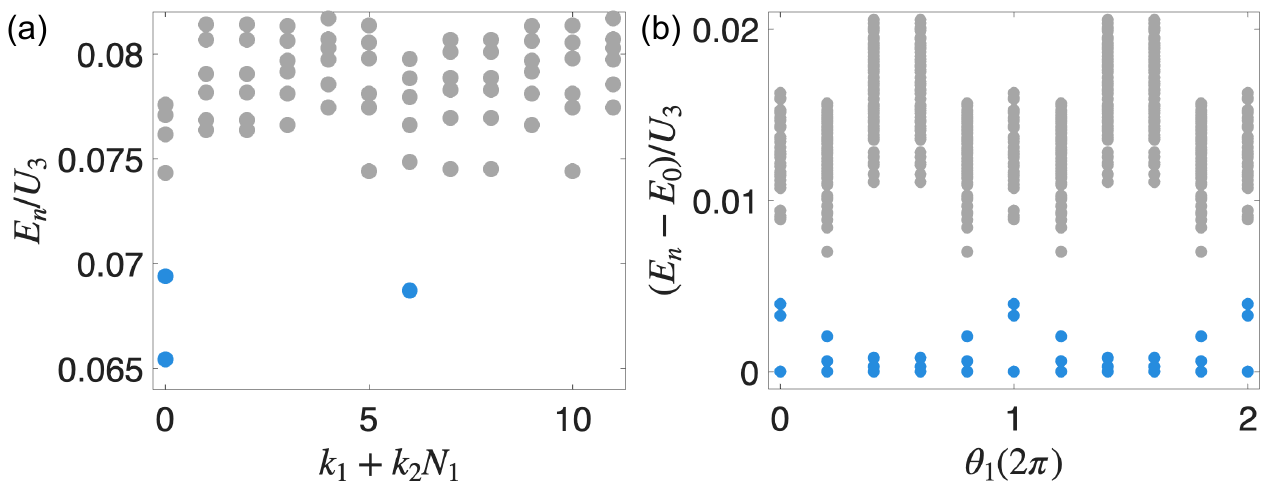}
	\caption{\textbf{Projected one-band ED spectra of the Moore-Read state in the gapped trivial flat band.} The results are computed by projecting the onsite three-body interaction to the gapped trivial band ($\eta=0.8$) at $\nu=1$ with $12$ soft-core bosons. $U_3$ is the strength of the three-body interaction that only sets the overall energy scale.
    (a) The characteristic 3-fold degenerate ground states appear in the momentum sectors expected from the generalized Pauli principle.
    (b) In the spectral flow, the three ground states remain well separated from the excited states upon flux insertion.
    }
	\label{figs_MR_projected_Nb12_eta0d8}
\end{figure}

Furthermore, we also perform the one-band ED simulation by projecting the onsite three-body interaction to the gapped trivial flat band ($\eta=0.8$).
This projection is well justified in the regime where the interaction strength of soft-core bosons is much smaller than the band gap.
Moreover, because the trivial band is exactly flat, the interaction strength in the projected simulations only sets the overall energy scale.
The resulting spectrum with 12 bosons is shown in Fig.~\ref{figs_MR_projected_Nb12_eta0d8}(a).
We find the characteristic 3-fold degenerate ground states of the Moore-Read phase, appearing in the momentum sectors expected from the generalized Pauli principle~\cite{Regnault2011_FCI}.
In addition, under twisted boundary conditions, the three ground states remain well separated from the excited states throughout flux insertion, as shown in Fig.~\ref{figs_MR_projected_Nb12_eta0d8}(b), which provides additional support for the identification of Moore-Read phase in the gapped trivial flat band.

\subsection*{C. Robustness of Laughlin and Moore-Read states in non-Chern flat bands}
\label{sec:robustness}
\begin{figure}[htp!]
	\centering
	\includegraphics[width=0.5\textwidth]{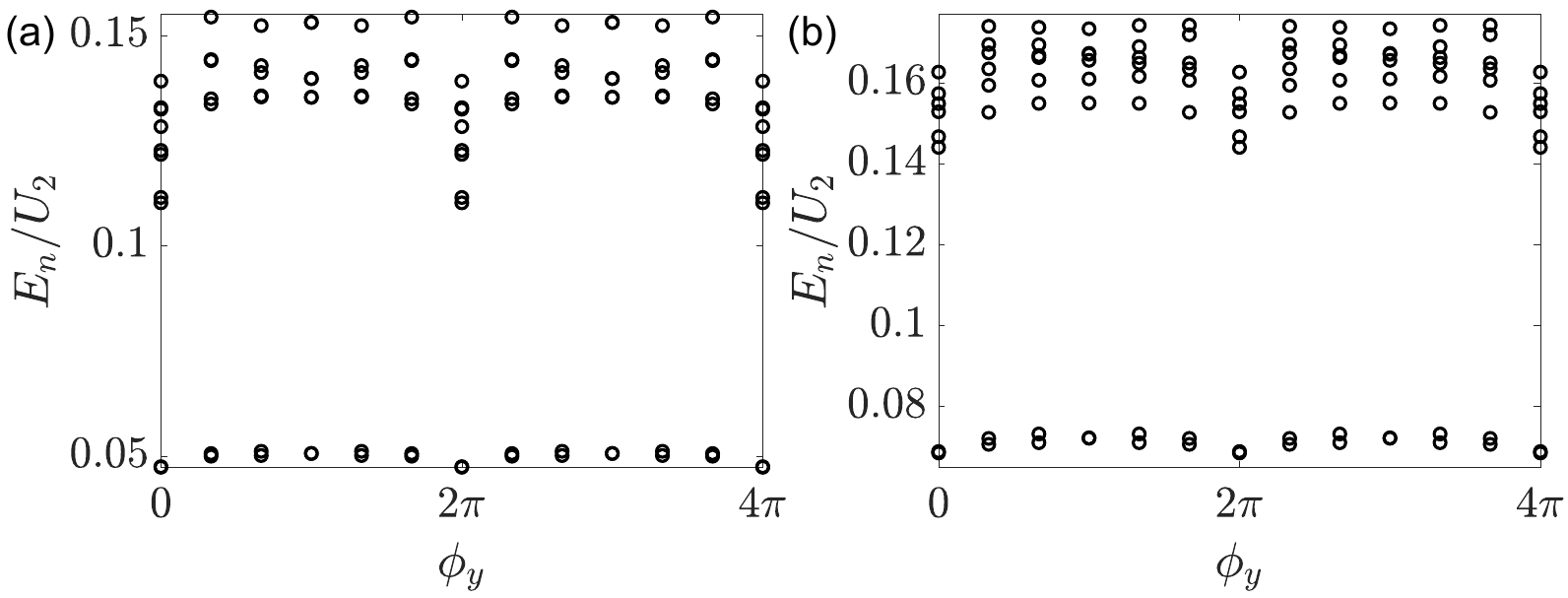}
	\caption{\textbf{Supplementary spectral flow of the Laughlin FCI in the gapless flat band from two-band ED simulations.} In Figs.~\ref{fig_laughlin_sfb}(a-b) of the main text, the spectral flow  of the degenerate ground states are shown in the insets. Here, we zoom out and add the low-energy excited states to further support the robust gap.
    The parameters are the same as Fig.~\ref{fig_laughlin_sfb}(a-b) of the main text with $\nu=1/2$ of the gapless flat band ($\eta=1$), and only very small $U_2=0.001$ interaction.
    Panels (a) and (b) are from $N=3\times4\times2$ and $N=4\times4\times2$ tori, respectively.
    }
	\label{figs_twist_sfb_softcore}
\end{figure}

In this section, we check the consistency of these FCIs at different system sizes and their existence in different circumferences.
First, in Fig.~\ref{fig_laughlin_sfb}(a-b) of the main text, we have shown the spectral flow  of the two degenerate ground states of the Laughlin state from two-band ED simulations in the insets. To further support the robust many-body gap upon flux insertion, we zoom out and and add the low-energy excited states as well in Fig. \ref{figs_twist_sfb_softcore}.

\begin{figure}[htp!]
	\centering
	\includegraphics[width=0.5\textwidth]{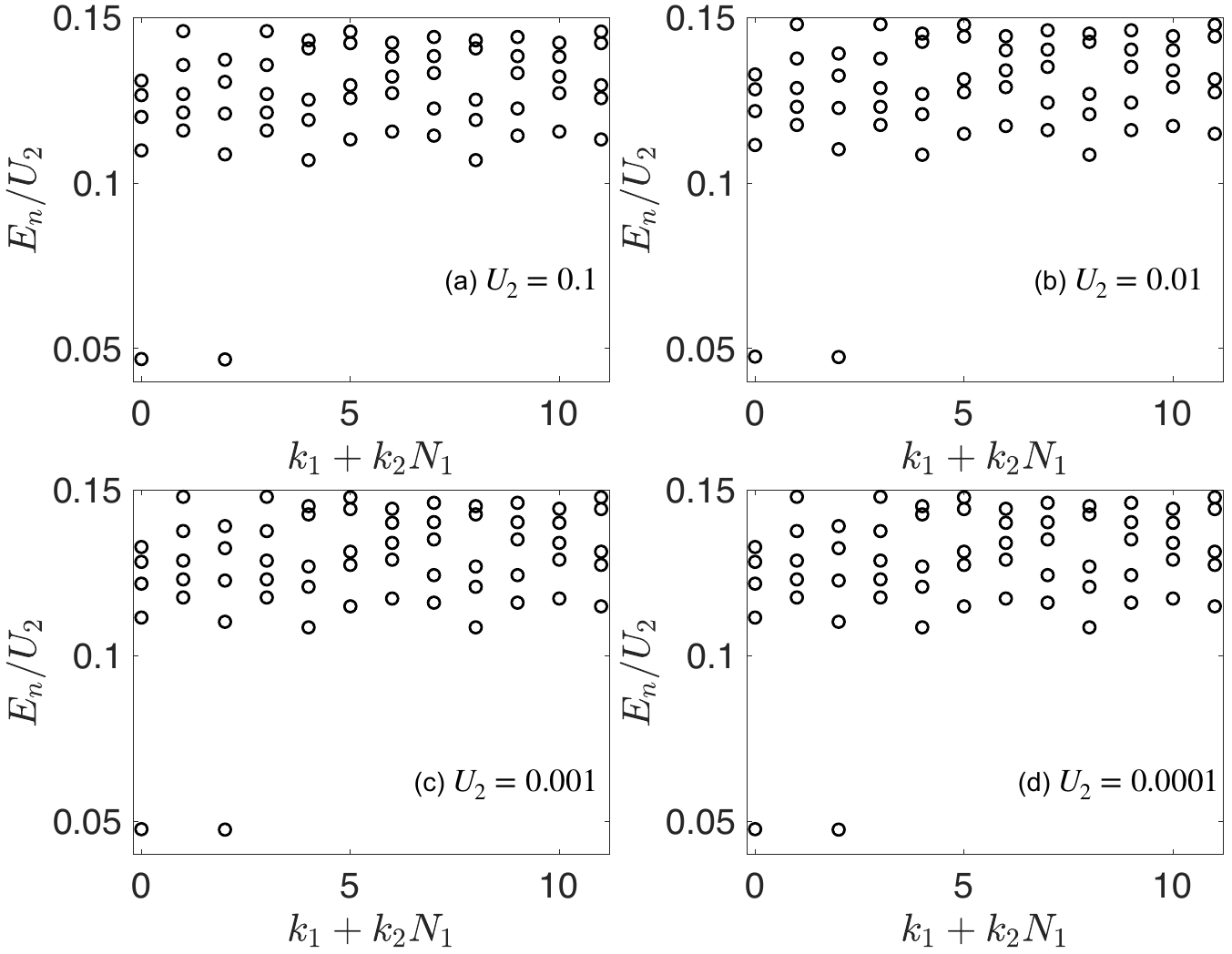}
	\caption{\textbf{Laughlin FCI in the gapless flat band with arbitrarily small onsite interactions.} We show the spectra of the FCI at $\nu=1/2$ of the gapless flat band with different $U_2$ from $N=3\times4\times2$ torus. Panel (c) has been showed in Fig.~\ref{fig_laughlin_sfb}(a) of the main text.
    }
	\label{figs_ed_different_U2_sfb}
\end{figure}

\begin{figure}[htp!]
	\centering
	\includegraphics[width=0.52\textwidth]{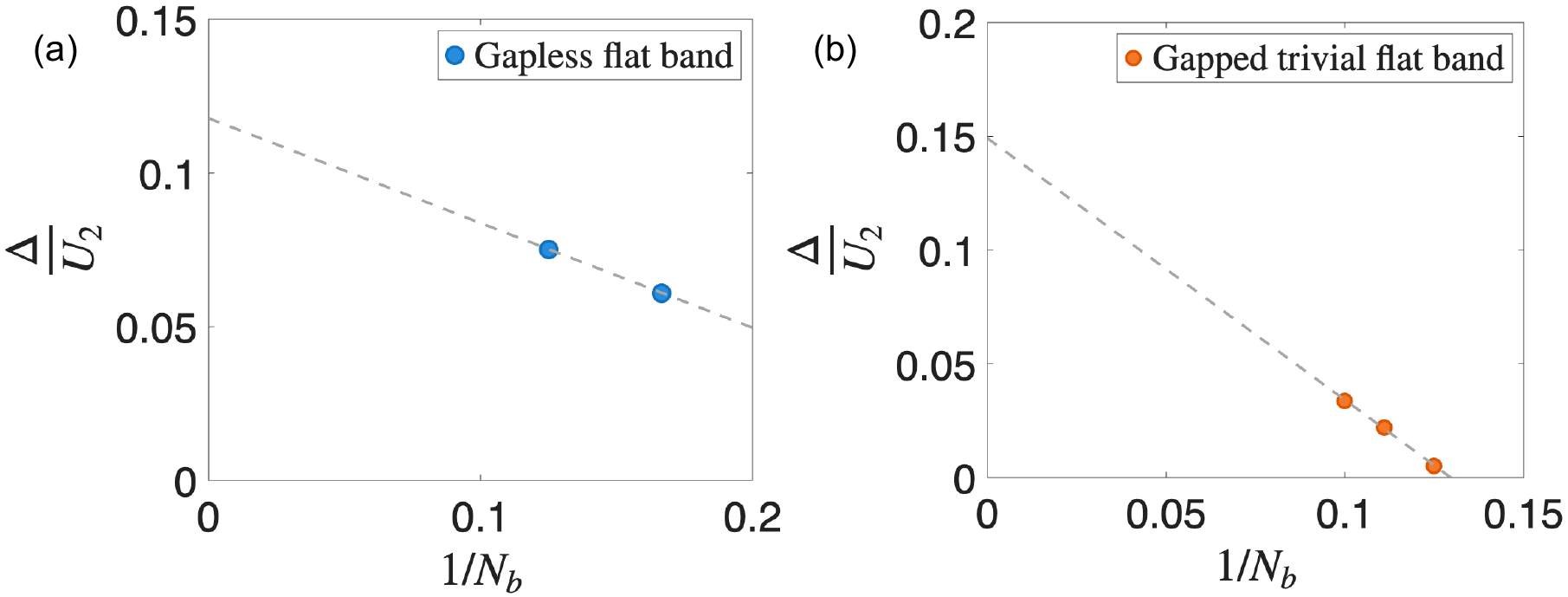}
	\caption{\textbf{Finite-size scaling of the many-body gaps for Laughlin states in non-Chern flat bands with soft-core bosons.}
    (a) The gaps in the gapless flat band from two-band ED simulations, and (b) the gaps in the gapped trivial flat band ($\eta=0.8$) from projected one-band ED simulations, are extrapolated in $1/N_b$.
    In both cases, the many-body gaps increase with the number of bosons.
    }
	\label{figs_extrapolation_gap_ed}
\end{figure}

In the main text, we have also mentioned that due to the exact flatness, arbitrarily small onsite interactions among the soft-core bosons can stabilize the FCI. In the main text, the example in the gapless flat band is taken as $U_2=0.001$.
To further support this, we show more two-band ED spectra with different $U_2$ in Fig.~\ref{figs_ed_different_U2_sfb}. It is clear that with the decreasing $U_2$, the spectra in units of $U_2$ almost converge, supporting that arbitrarily small $U_2$ could stabilize this FCI at $\nu=1/2$ of the gapless flat band.
Similarly for the gapped trivial flat band, arbitrarily small onsite interactions could induce the FCIs, which is supported by the projected one-band ED simulations. The example of the Moore-Read state is shown in Fig.~\ref{figs_MR_projected_Nb12_eta0d8} and that of the Laughlin states will be discussed later in Fig.~\ref{figs_extrapolation_gap_ed}.

\begin{figure}[htp!]
	\centering
	\includegraphics[width=0.5\textwidth]{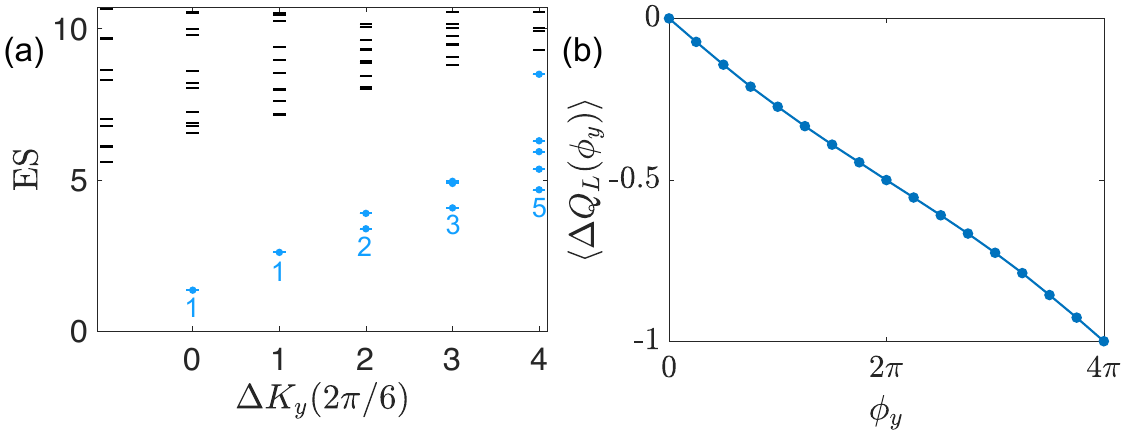}
	\caption{\textbf{Robust FCI with strong NN interaction.} Here, we consider the Laughlin state in the gapless flat band at $\nu=1/2$ filling with hard-core bosons, and $V_1=9$ is taken for these results. (a) The momentum-resolved ES. (b) The charge pumping result.
    }
	\label{figs_Laughlin_sfb_V1}
\end{figure}

Throughout this work, we mainly consider the onsite interactions. Here, in the gapless flat band model, we take the hard-core boson case for example to explore the NN (nearest neighbor) repulsion: $H_V=V_1\sum_{\langle ij \rangle}n_in_j$.
At $\nu=1/2$, we find that the FCI is stable even when $V_1$ is larger than the total width of the two bands ($W=9$). We show the iDMRG results with $V_1=9$ in Fig.\ref{figs_Laughlin_sfb_V1}.
The characteristic edge-mode counting from the ES and the charge pumping with one boson pumped after adiabatically inserting $4\pi$ fluxes confirm the robustness of the FCI with very large $V_1$ interactions and suggest that there might be no upper limit of $V_1$ for this FCI.

Next, we discuss the finite-size scaling of these FCIs for consistency checks of their robustness.
For the Laughlin state in the gapless flat band with soft-core bosons, using the two-band ED spectra for $N_b=6,\ 8$ bosons in Fig.~\ref{fig_laughlin_sfb} (a-b) of the main text, we find that the many-body gap increases with boson number. A linear extrapolation in $1/N_b$ yields a value of approximately $\frac{\Delta}{U_2}\sim0.118$, as shown in Fig.~\ref{figs_extrapolation_gap_ed} (a).
For the Laughlin state in the gapped trivial band ($\eta=0.8$) with soft-core bosons, we similarly extrapolate the many-body gap from projected one-band ED results for $N_b=8,9,10$ bosons, as shown in Fig.~\ref{figs_extrapolation_gap_ed} (b).
The gap again increases with system size, and the extrapolated value is $\frac{\Delta}{U_2}\sim 0.149$.
In both cases, the many-body gaps increase with the number of bosons.

\begin{figure}[htp!]
	\centering
	\includegraphics[width=0.52\textwidth]{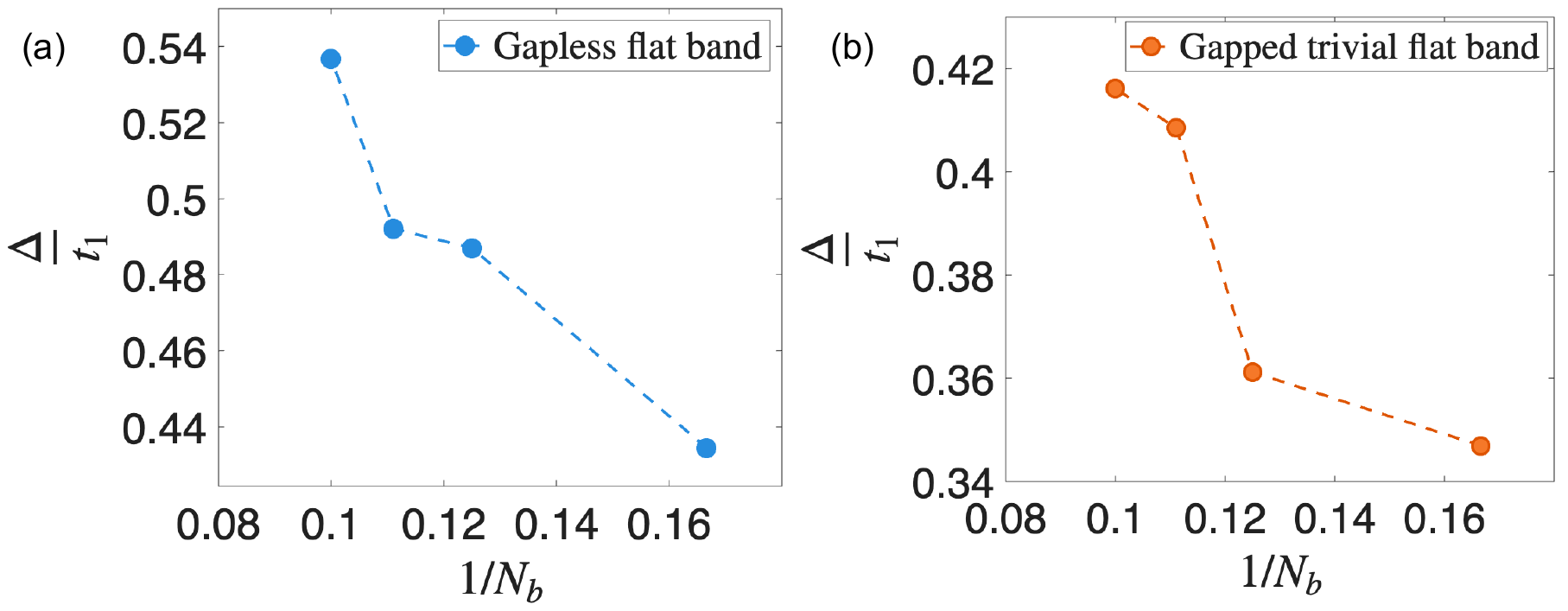}
	\caption{\textbf{Finite-size analysis of ED simulations with hard-core bosons for the Laughlin FCI.}
    We show the results for (a) the gapless flat band and (b) gapped trivial flat band (at $\eta=0.8$).
    The energy gaps are shown in units of hopping amplitude $t_1$.
    In both cases, the energy gaps are increasing with the system sizes.
    }
	\label{figs_ED_gap_hardcore_Laughlin}
\end{figure}

\begin{figure}[htp!]
	\centering
	\includegraphics[width=0.52\textwidth]{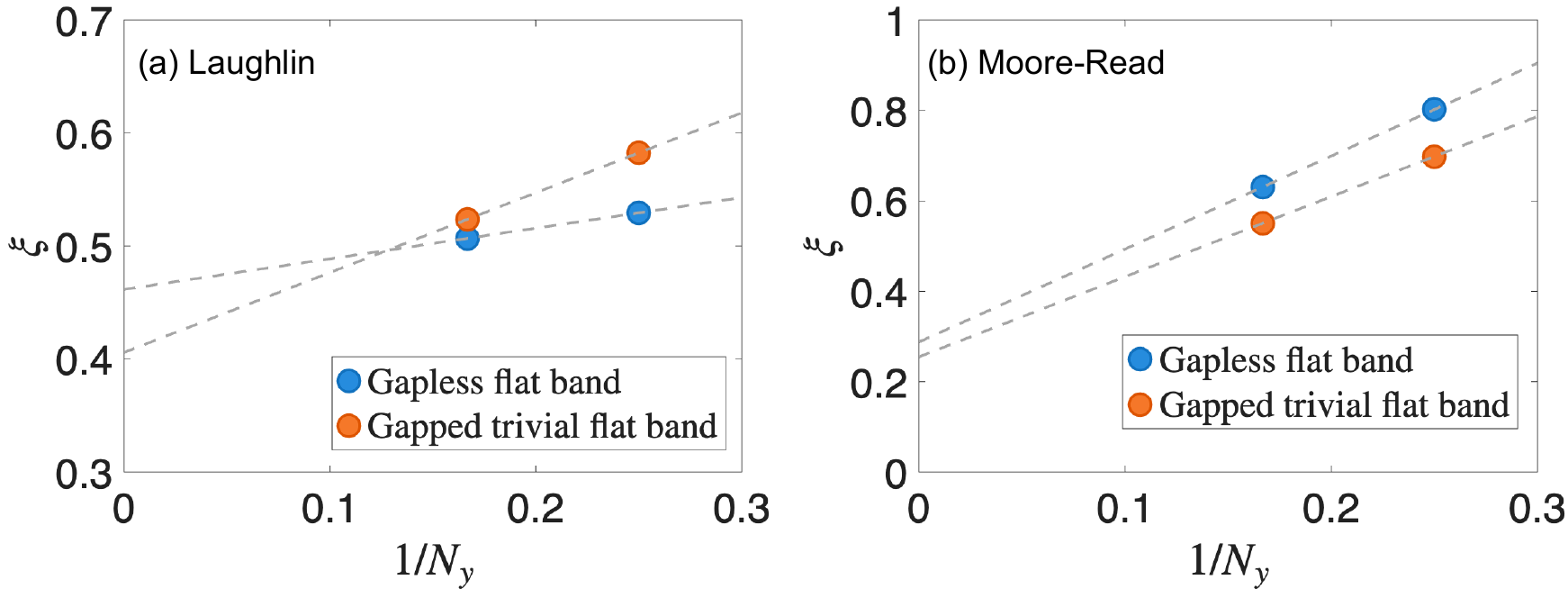}
	\caption{\textbf{Finite-size scaling of the correlation length for Laughlin and Moore-Read states in non-Chern flat bands with hard-core bosons.}
    For both the (a) Laughlin at $\nu=1/2$ and (b) the Moore-Read states at $\nu=1$, the correlation lengths from iDMRG simulations decrease when increasing the circumference $N_y$ of the cylinder. We do the extrapolations in $1/N_y$.
    }
	\label{figs_extrapolation_correlation}
\end{figure}

Similarly, for the Laughlin states in the gapless flat band and in the gapped trivial band ($\eta=0.8$) with hard-core bosons, we assess the finite-size scaling using both ED simulations of energy gaps and iDMRG simulations by examining the correlation length as a function of cylinder circumference (in iDMRG simulations, the inverse correlation length can be interpreted as the excitation gap).
For the ED results shown in Fig.~\ref{figs_ED_gap_hardcore_Laughlin}, for both the gapless flat band and the gapped trivial flat band, the energy gaps are increasing with the system sizes.
For the iDMRG results shown in Fig.~\ref{figs_extrapolation_correlation} (a), the correlation length decreases as the cylinder circumference increases, and the extrapolated values in $1/N_y$ are approximately $0.46$ and $0.41$, respectively.
These finite and short correlation lengths, both smaller than one lattice spacing, provide strong evidence that the corresponding many-body gaps  remain robust consistently in these accessible system sizes.
In addition, we perform the same analysis for the Moore-Read states in the gapless flat band and in the gapped trivial band ($\eta=0.8$) with hard-core bosons.
As shown in Fig.~\ref{figs_extrapolation_correlation} (b), the correlation lengths again decrease with increasing cylinder circumference, and the extrapolated values are approximately $0.29$ and $0.26$, respectively.
In summary, these consistency checks support that these FCIs are robust in the currently accessible system sizes. We also note that larger-scale calculations, possibly with more advanced numerical methods, would be beneficial for a controlled extrapolation to the thermodynamic limit.

\begin{figure}[htp!]
	\centering
	\includegraphics[width=0.6\textwidth]{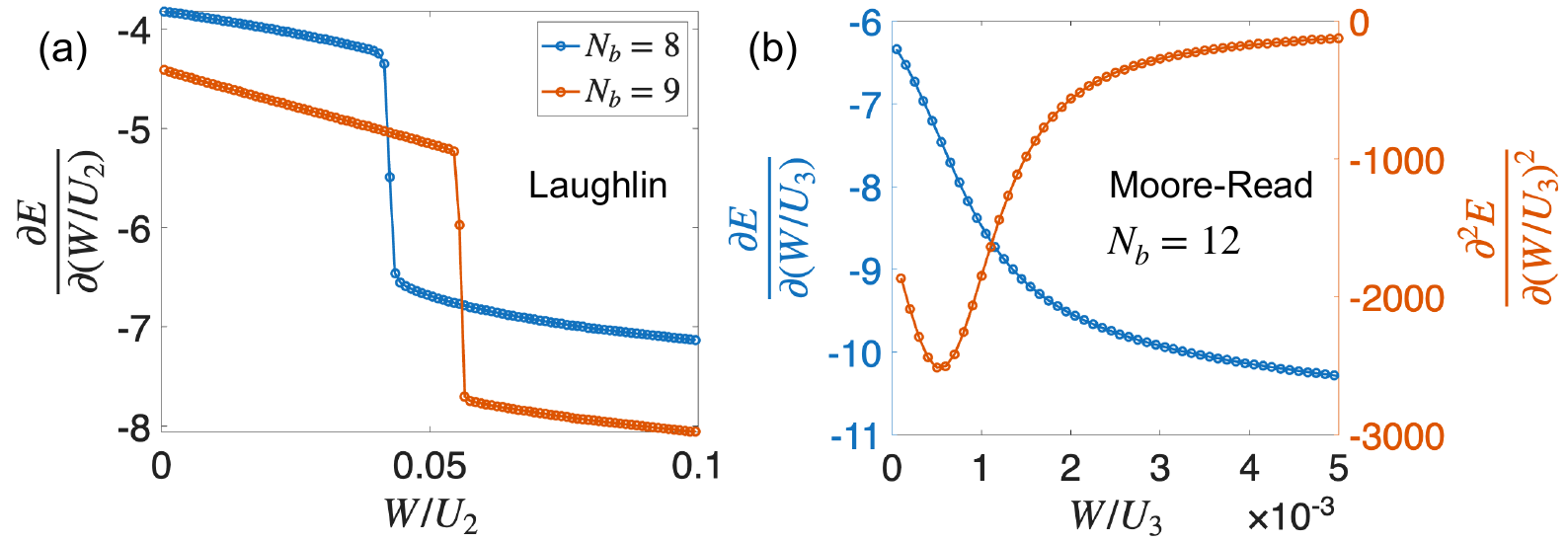}
	\caption{\textbf{FCIs in non-Chern bands against bandwidth.}
    We obtain the results in projected one-band ED simulations for the gapped trivial flat band ($\eta=0.8$).
    (a) For the Laughlin state at $\nu=1/2$, we show the first derivative of energy with respect to the ratio between the the bandwidth ($W$) and the strength of two-body onsite interaction ($U_2$)  in the $N_b=8$ and $N_b=9$ systems.
    (b) For the Moore-Read state at $\nu=1$, we show the first and second derivatives of energy with respect to the ratio between the bandwidth ($W$) and the strength of three-body onsite interaction ($U_3$)  in the $N_b=12$ system.
    }
	\label{figs_FCI_SF_dispersion}
\end{figure}

Apart from the robustness of many-body gaps of the FCIs at different system sizes, another interesting aspect is the robustness of these states against bandwidth, as these non-Chern bands are always exactly flat and stronger dispersion might introduce additional instability~\cite{Lu2025_FCI_SF}.
We first consider $\eta=0.8$ for example and project the onsite interactions to the gapped trivial flat band for one-band ED simulations, where
a controllable dispersion can be straightforwardly introduced while keeping the underlying single-particle wave functions and quantum geometry of the non-Chern band unchanged.
We add a hexagonal periodic potential to the projected Hamiltonian: $H=H_\mathrm{proj}+E(\mathbf{k})$, with $E(\mathbf{k})=-\frac{W}{3}\sqrt{1+4\cos\!\left(\frac{\sqrt3 k_x}{2}\right)\cos\!\left(\frac{k_y}{2}\right)
+4\cos^2\!\left(\frac{k_y}{2}\right)}$, where $W$ denotes the induced bandwidth.
For the Laughlin state at $\nu=1/2$, we gradually tune the ratio between the added bandwidth and the strength of the projected onsite two-body interaction: $\frac{W}{U_2}$ from $0$. When $W$ is large enough, the ground state is the superfluid phase.
As shown in Fig.~\ref{figs_FCI_SF_dispersion} (a), the first-derivative of the ground-state energy with respect to $\frac{W}{U_2}$ for the $N_b=8$ and $N_b=9$ (number of bosons) systems exhibits sharp discontinuity at $\frac{W}{U_2}\sim0.043,\ 0.056$, respectively, indicating the first-order FCI-to-superfluid transitions.
These results suggest that the Laughlin states remain robust against finite dispersions and the robustness appears to increase with the system size.

Furthermore, we show the results of the Moore-Read state with $N_b=12$ in Fig.~\ref{figs_FCI_SF_dispersion} (b).
The first-derivative of the ground-state energy with respect to the ratio between the added bandwidth and the strength of the projected onsite three-body interaction, $\frac{W}{U_3}$, evolves continuously, while the peak in the second derivative identifies the transition point. The nature of this transition is difficult to determine from this finite size calculation limited by the computation capacity.
Importantly, this Moore-Read state can also exist against small dispersions but the transition point to the superfluid phase is $\frac{W}{U_3}\sim0.0005$, suggesting the Moore-Read state of soft-core bosons in the non-Chern band is more fragile against band dispersion than the Laughlin states.

\begin{figure}[htp!]
	\centering
	\includegraphics[width=0.72\textwidth]{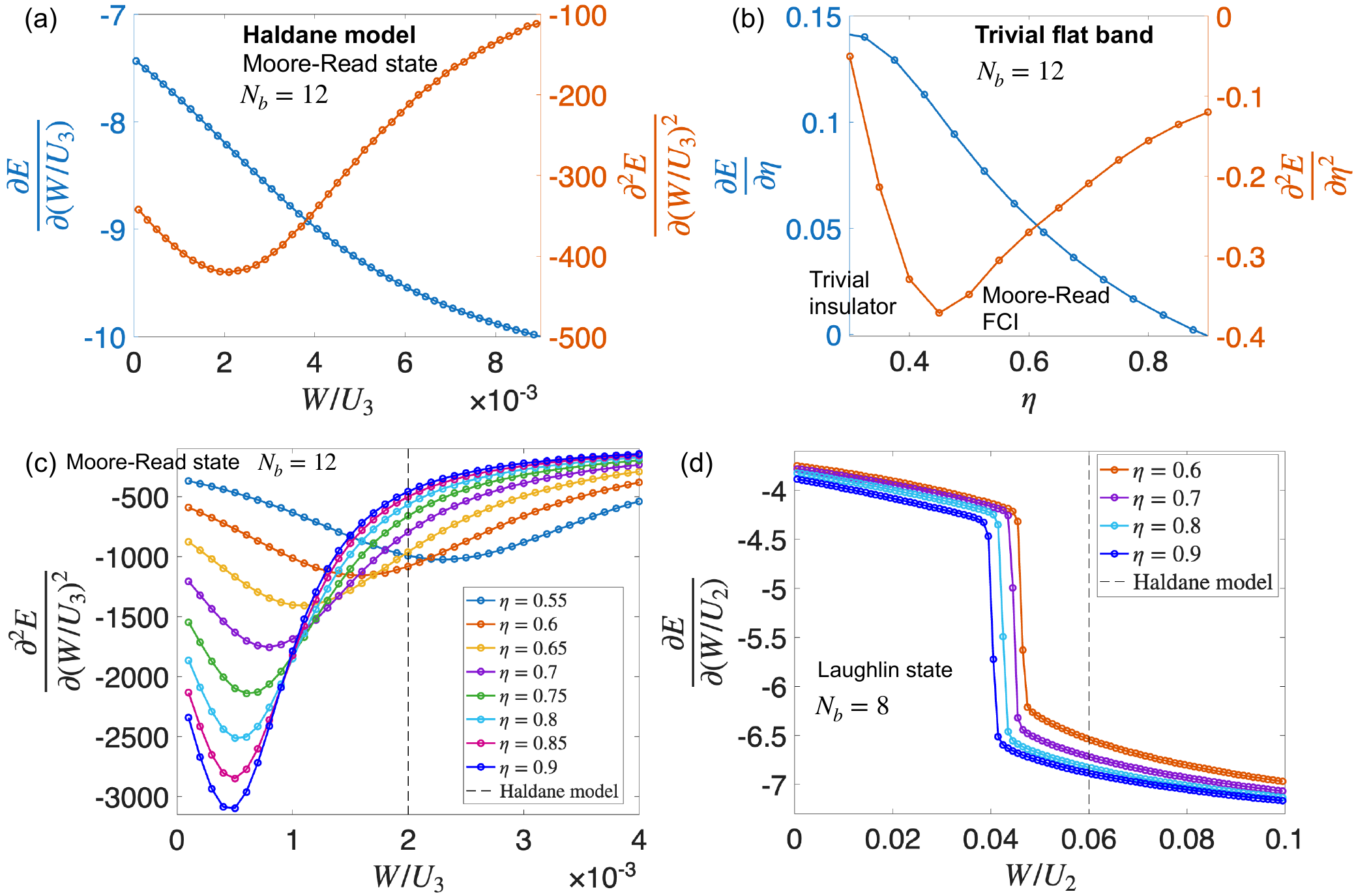}
	\caption{\textbf{$\eta$-dependence and benchmark of the bandwidth tolerance.}
    In panel (a), the pure three-body onsite interaction is projected to the nearly flat Chern band of the Haldane model with almost optimal single-particle parameters, and the same one-body periodic potential is applied for checking the robustness of Moore-Read state with $N_b=12$ bosons against increasing bandwidth. According to the derivatives of ground-state energy, the transition to the superfluid phase occurs at $W/U_3\sim2\times10^{-3}$.
    In panel (b), we show that in the projected simulation with three-body onsite interaction and $N_b=12$, when decreasing $\eta$, the transition from the Moore-Read state to the trivial insulator occurs at $\eta\sim0.45$ according to the derivatives of ground-state energy.
    This determines the parameter range over which the bandwidth tolerance of the Moore–Read state can be evaluated.
    The results are shown in panel (c). The peaks of second derivative of ground-state energy with $N_b=12$ suggest that the critical bandwidth is increasing when $\eta$ is decreasing. More importantly, the critical bandwidth of the Moore-Read state in the trivial flat band could be very close to and even surpass that of the Moore-Read state in the Haldane model (dashed line), suggesting the robustness of such a non-Abelian FCI in the non-Chern flat-band system.
    In panel (d), we show the $\eta-$dependence of the Laughlin state in the non-Chern flat band with $N_b=8$ and projected onsite two-body interaction. Similarly, the critical bandwidth is increasing when $\eta$ gets smaller and is very close to that of the Laughlin state in the Haldane model (dashed line).
    }
	\label{figs_bandwidth_eta}
\end{figure}

To further understand the bandwidth tolerance, especially for the Moore-Read FCI, we consider an optimized Haldane model with a nearly flat lower Chern band and project the same onsite three-body interaction into this band.
For the non-interacting part of the Haldane model ($H_t=-\sum_{i,j}t_{ij}c_i^\dagger c_j^{\ }+h.c.$ with parameters $t_1$, $t_2e^{\pm i\phi}$, and $t_3$), the optimized parameters are taken from \cite{chen2026_fermion_fci_ipeps}: $t_2/t_1=0.7$, $t_3/t_1=-0.9$, and $\phi=0.35\pi$.
We then apply the same hexagonal periodic potential used in our non-Chern flat-band calculation to tune the bandwidth $W$, and the derivatives of the ground-state energy for $N_b=12$ in Fig.~\ref{figs_bandwidth_eta} (a) show that the transition point to the superfluid phase is around $W/U_3\sim2\times10^{-3}$.
This value is only a few times larger than the value found in the non-Chern flat band at $\eta=0.8$, showing that the small critical bandwidth is not a disadvantage of the non-Chern flat-band system in comparison with an optimized Chern band.

We further examine the effect of tuning $\eta$ on the robustness of the Moore-Read state in our non-Chern band system against bandwidth.
In Fig.~\ref{figs_bandwidth_eta} (b), we show that, when decreasing $\eta$, the transition point to the trivial insulator in the projected simulation is $\eta\sim0.45$ according to the derivatives of ground-state energy.
Next, we examine the bandwidth tolerance within the Moore-Read phase regime when tuning $\eta$.
As shown in Fig.~\ref{figs_bandwidth_eta} (c), when decreasing $\eta$,  the robustness against finite bandwidth is actually increasing.
More importantly, the critical bandwidth remains comparable to, or even slightly exceed that of the optimized Haldane model.

Such analysis is also done for the Laughlin state when tuning $\eta$, and similarly, the robustness of the FCI in the non-Chern band against bandwidth is increasing when decreasing $\eta$ and the critical value is at the same order and very close to that of Laughlin state in the optimized Haldane model, as shown in Fig.~\ref{figs_bandwidth_eta} (d).
We also note in the literature that the bandwidth-induced transition from the Laughlin state to the superfluid state is both theoretically analyzed and numerically explored,  for example by adding periodic potentials in a projected Landau-level setting~\cite{wang2026emergentqed3bosoniclaughlin}, where the critical bandwidth is also of order ($10^{-2}$) in units of the projected two-body interaction strength, comparable to those of Laughlin states for the non-Chern-band here and Haldane-model cases.
In contrast, an analogous controlled analysis for Moore–Read states remains much less explored. This makes the finite-bandwidth stability of the Moore–Read phase an important issue in its own right, and provides further motivation for our comparison with an optimized Haldane Chern-band benchmark.

\subsection*{D. Quantum phase diagram by tuning \texorpdfstring{$\eta$}{eta}}
\label{sec:phase-diagram}
\begin{figure}[htp!]
	\centering
	\includegraphics[width=0.6\textwidth]{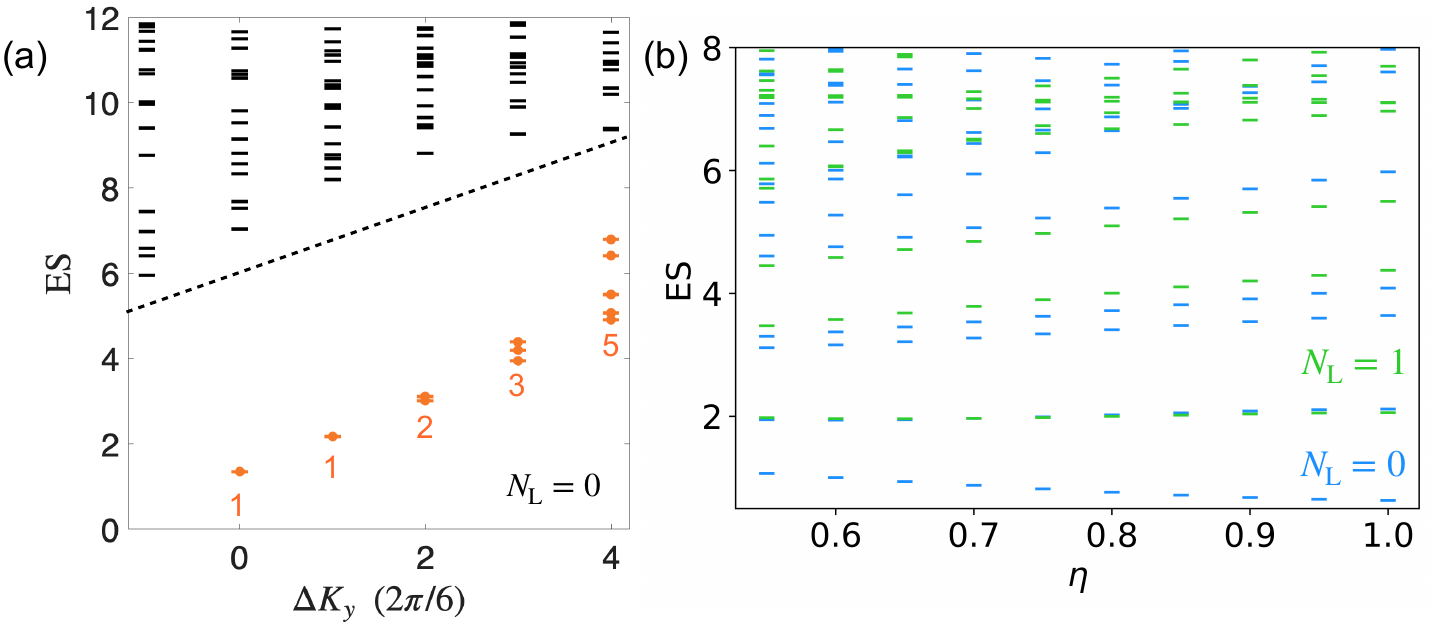}
	\caption{\textbf{Laughlin state in the gapped trivial flat band ($\eta=0.8$) with hard-core bosons.}
    (a) The momentum-resolved entanglement spectrum from iDMRG simulations at $\nu=1/2$, and
    the characteristic edge-mode counting $\{1,1,2,3,5,...\}$ is in agreement with the edge conformal field theory, supporting the Laughlin ground state.
    (b) The continuous and smooth evolution of entanglement spectra (of the two lowest charge sectors) for the Laughlin states in the gapless flat band and in the gapped trivial flat band with  hard-core bosons upon tuning $\eta$, suggesting the adiabatic connections of the many-body state.
    }
	\label{figs_Laughlin_eta0d8_hardcore}
\end{figure}

For the gapped trivial flat band, we have shown that arbitrarily small onsite $U_2$ interactions could stabilize the Laughlin state.
Here, we show that the Laughlin state in the gapped trivial flat band is also stable with infinitely large onsite interaction, as supported by the momentum-resolved entanglement spectrum at $\nu=1/2$ filling of the gapped trivial flat band ($\eta=0.8$) with hard-core bosons [Fig.~\ref{figs_Laughlin_eta0d8_hardcore} (a)].
Besides, similar to the Moore-Read states, this Laughlin state is also adiabatically connected to that in the gapless flat band, as shown in Fig.~\ref{figs_Laughlin_eta0d8_hardcore} (b).

\begin{figure}[htp!]
	\centering
	\includegraphics[width=0.52\textwidth]{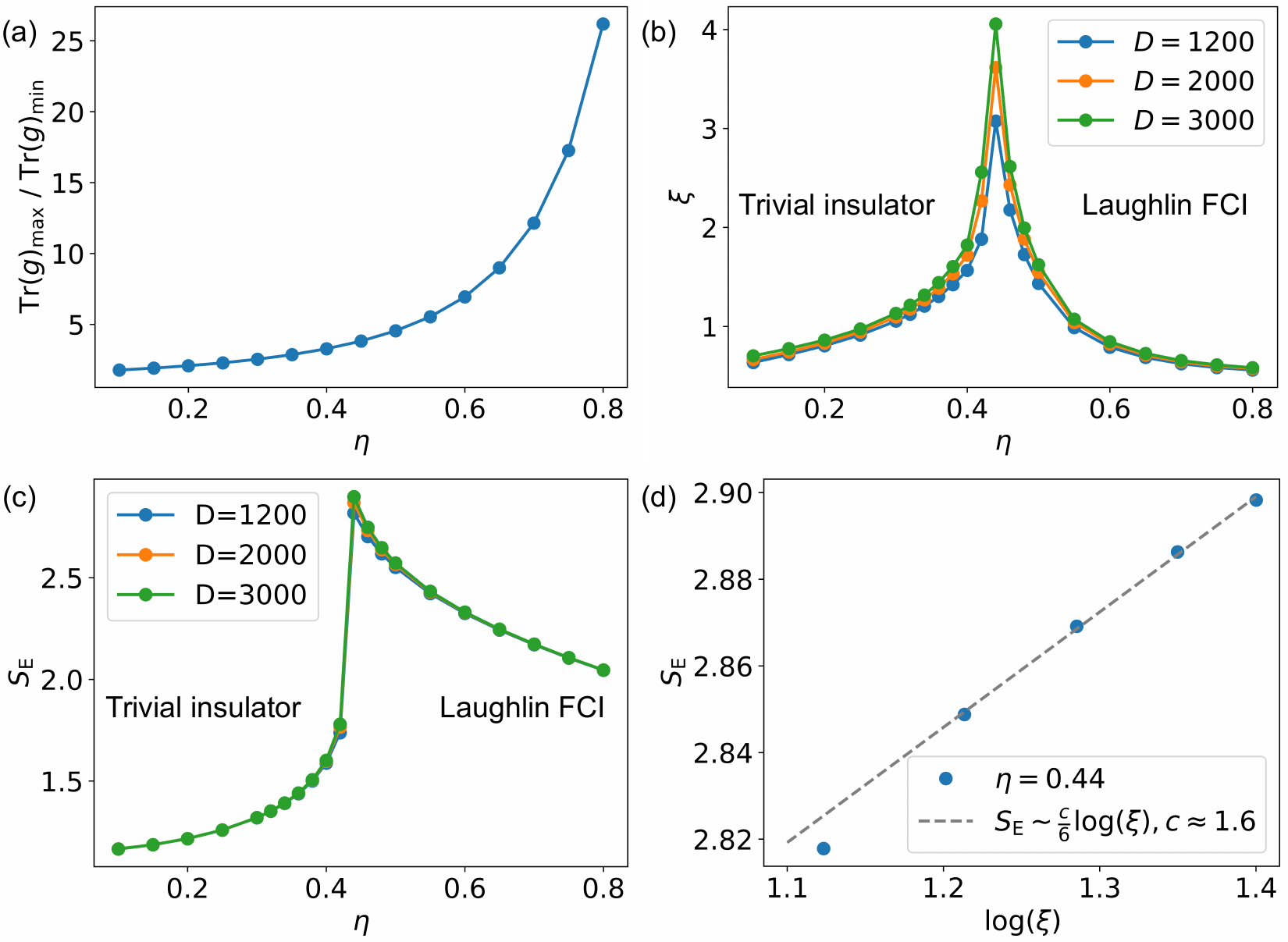}
	\caption{\textbf{Laughlin-state-to-trivial-insulator transition by tuning $\eta$.} (a) The ratio between the maximum and minimum values of the trace of quantum metric, as an indicator of the sharpness of the quantum geometry peak at $\Gamma$. The smaller the $\eta$, the less sharper the quantum geometry peak at $\Gamma$.
    (b-d) panels are iDMRG results on the circumference $N_y=4$ cylinders at $\nu=1/2$ with hard-core bosons, including (b) the correlation length and (c) the bipartite entanglement entropy as functions of $\eta$, and (d) the fitting of central charge at the critical point $\eta\sim0.44$.
    }
	\label{figs_transition_laughlin_trivial_insulator}
\end{figure}

At relatively large $\eta$ where the FCIs are robust, the quantum geometry of the gapped trivial flat band exhibits a sharp peak near $\Gamma$.
Upon decreasing $\eta$, the peak of quantum geometry becomes less sharp [Fig.~\ref{figs_transition_laughlin_trivial_insulator} (a)], and the Laughlin FCI continuously loses stability and eventually gives way to a topologically trivial insulator.

Regarding the nature of the transition, our iDMRG results are consistent with a continuous one. In particular, the correlation length and bipartite entanglement entropy grows strongly and appears to diverge near the critical point $\eta\sim0.44$, as shown in Fig.~\ref{figs_transition_laughlin_trivial_insulator} (b,c).
Moreover, our iDMRG simulations allow us  to fit the bipartite entanglement entropy as a function of the correlation length at the critical point~\cite{Scott2016_CFL}, as shown in Fig.~\ref{figs_transition_laughlin_trivial_insulator} (d), and we obtain a finite central charge, further supporting the presence of critical behavior, which might offer a new and useful setting for exploring continuous topological transitions beyond the conventional Ginzburg-Landau paradigm~\cite{Wen1993_fqh_insulator_transition, Barkeshli2014_fci_sf,Song2023_deconfined_criticality_duality}.
This also provides a novel example of quantum-geometry-driven topological many-body transition without changing band topology, dispersion, or interaction strength, while the latter ones are the common tuning forces in exploring FCIs in Chern-band systems.

\subsection*{E. Two-body interactions at \texorpdfstring{$\nu=1$}{nu=1}}
\label{sec:two-body}

\begin{figure}[htp!]
	\centering
	\includegraphics[width=0.8\textwidth]{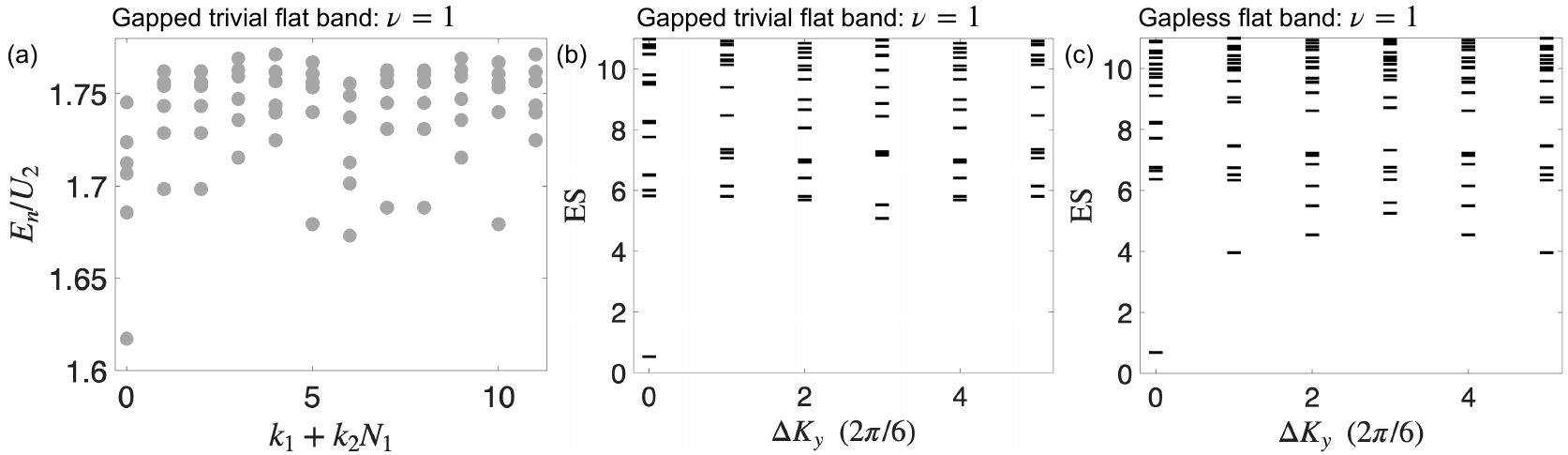}
	\caption{\textbf{Results of two-body interactions at $\nu=1$ of non-Chern flat bands.} (a) One-band ED spectrum by projecting the two-body onsite interaction to the gapped trivial flat band ($\eta=0.8$) with $N_b=12$.
    The momentum-resolved entanglement spectra with (two-body) hard-core bosons from iDMRG simulations of the (b) gapped trivial flat band ($\eta=0.8$) and (c) gapless flat band. No signature of Moore-Read state is identified.
    }
	\label{figs_nu_1_twobody_onsite}
\end{figure}

In this work, the Moore-Read states in non-Chern flat bands are realized by onsite three-body interactions, while we also note the theoretical study reporting the bosonic Moore-Read state arising from purely two-body onsite interactions in Chern-band systems~\cite{Palm2021_bosonic_pfaffian}.
Therefore, we have also examined the $\nu=1$ filling of the present non-Chern bands with purely two-body interactions, for both soft-core and hard-core bosons.
For the soft-core boson case, we project the onsite two-body interaction to the gapped trivial band ($\eta=0.8$) and the ED spectrum for $N_b=12$ bosons is shown in Fig.~\ref{figs_nu_1_twobody_onsite}(a).
However, in contrast to the projected three-body-interaction case [Fig.~\ref{figs_MR_projected_Nb12_eta0d8}], we do not find evidence for the expected 3-fold nearly degenerate ground states of the Moore-Read phase.
For the two-body hard-core boson case ($n_\mathrm{max}=1$), we further examine the momentum-resolved entanglement spectra for both the gapless-flat-band and gapped-trivial-flat-band cases from iDMRG simulations, as shown in Fig.~\ref{figs_nu_1_twobody_onsite}(b,c), where we do not observe the characteristic counting of the edge conformal field theory in either case.
However, we cannot exclude the possibility that the non-abelian FCI may exist in some other narrow parameter window.

\subsection*{F. Quantum-geometry adaptiveness}
\label{sec:geometry-adaptiveness}

\begin{figure}[htp!]
	\centering
	\includegraphics[width=0.3\textwidth]{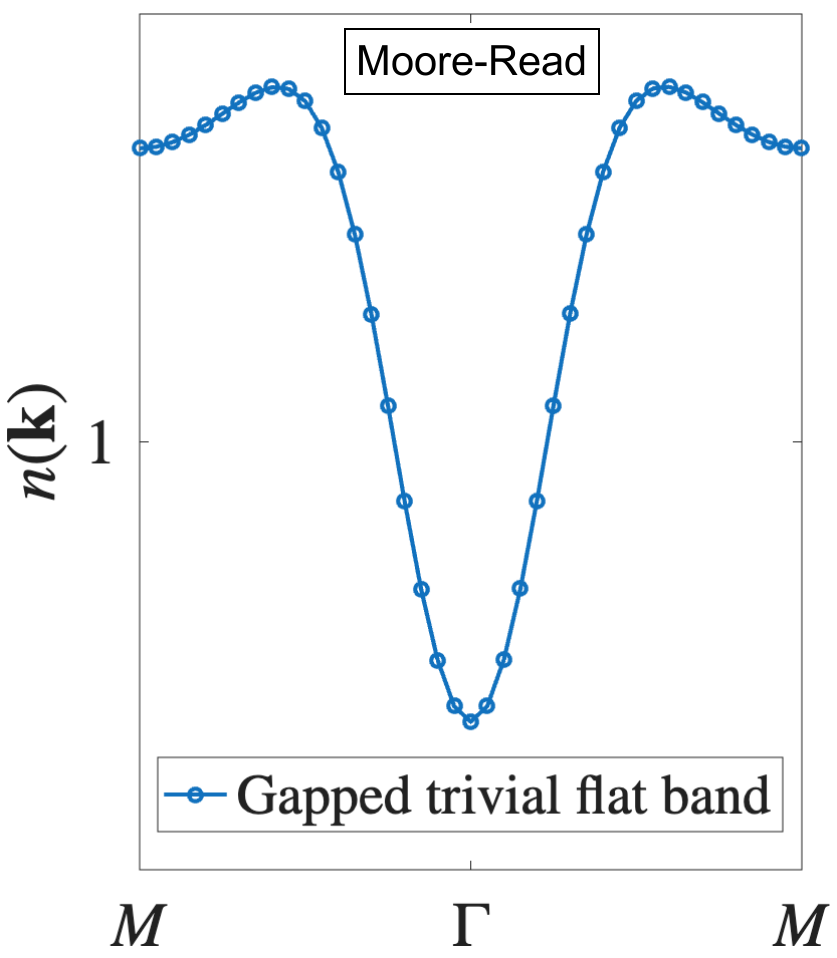}
	\caption{\textbf{ Boson occupation $n(\mathbf{k})$ for Moore-Read FCI in the gapped trivial flat band ($\eta=0.8$).}
    This is from iDMRG results with hard-core bosons on cylinders of circumference $N_1=6$. The boson occupation is plotted on the same path as that for the Laughlin state in Fig.~\ref{fig_laughlin_trivial_band}(a).
    Similarly, the boson occupation at $\Gamma$ (where the flat-band quantum geometry peaks sharply) is strongly suppressed.
    }
	\label{figs_nk_MR_eta0d8_idmrg}
\end{figure}

In the main text, we have introduced that in the gapped trivial flat band, due to the sharp peak of quantum metric at $\Gamma$, the boson occupation of the Laughlin state is inhomogeneous and strongly suppressed at $\Gamma$. Here, we show the similar effect observed for the Moore-Read FCI, with spontaneously developed inhomogeneous
distributions to occupy regions with small quantum metric, and the iDMRG result with hard-core bosons is shown in Fig.~\ref{figs_nk_MR_eta0d8_idmrg}.

\begin{figure}[htp!]
	\centering
	\includegraphics[width=0.65\textwidth]{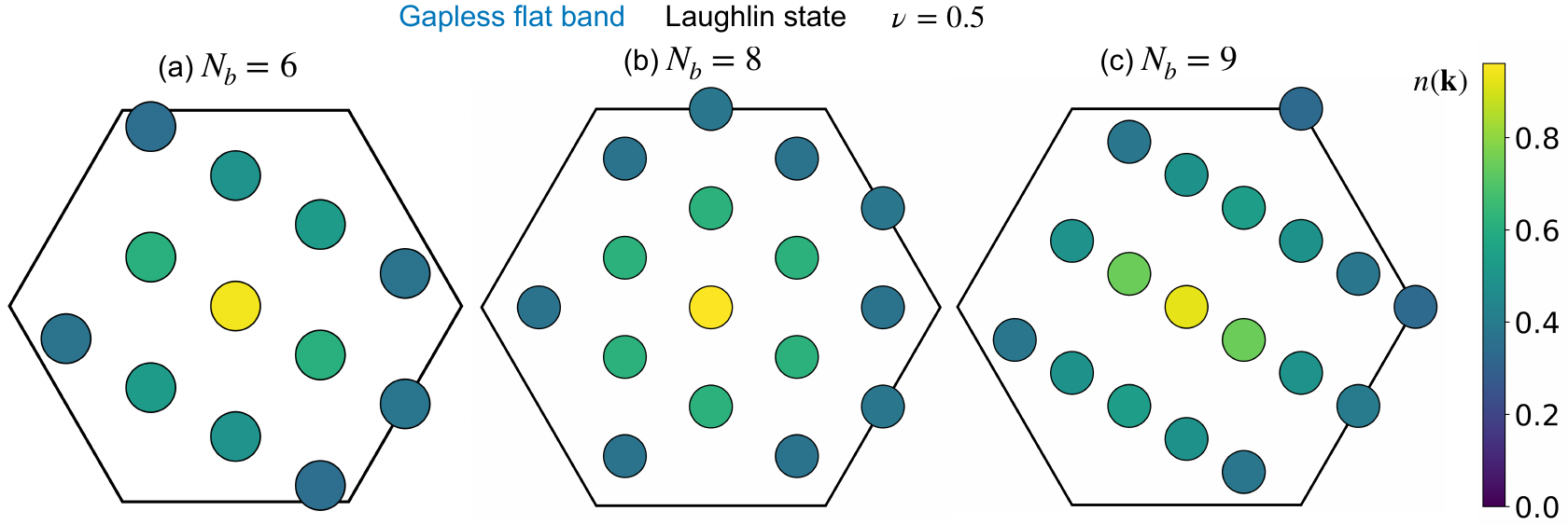}
	\caption{\textbf{Consistency checks of $n(\mathbf{k})$ for Laughlin FCI in the gapless flat band.}
    The momentum-space boson occupations of the Laughlin state in the gapless flat-band are shown from different sizes of full ED simulations with hard-core bosons.
    The boson occupations consistently peaks at the band-touching point.
    }
	\label{figs_nk_hc_Laughlin_ED}
\end{figure}

\begin{figure}[htp!]
	\centering
	\includegraphics[width=0.5\textwidth]{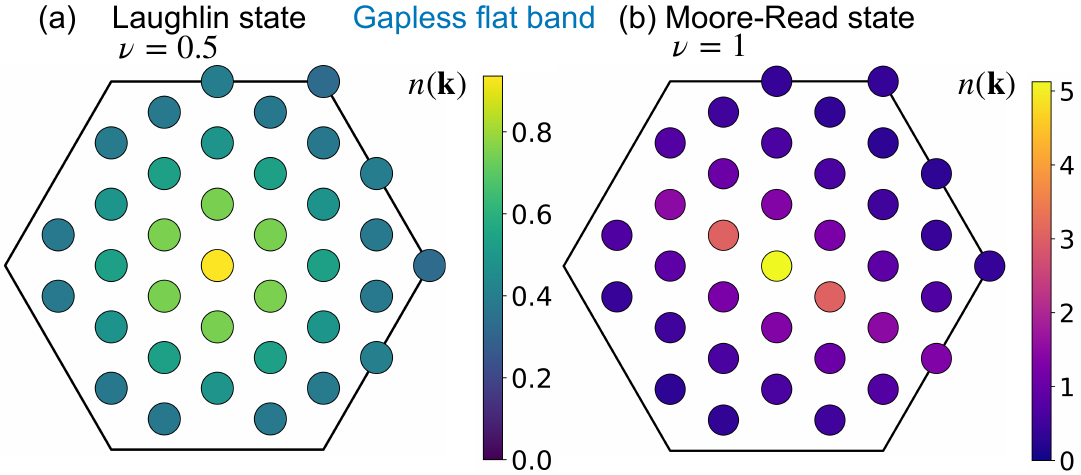}
	\caption{\textbf{Consistency checks of  $n(\mathbf{k})$ from iDMRG simulations with hard-core bosons.}
    These results are obtained in circumferences $N_1=6$ cylinders. Consistenly, we observe the peaks of $n(\mathbf{k})$ at the band-touching point for both the Laughlin and Moore-Read FCIs and it is even more prominent for the Moore-Read phase.
    }
	\label{figs_nk_hc_idmrg}
\end{figure}

In the main text, we have also shown that the $n(\mathbf{k})$ of the FCIs in the gapless flat band would adaptively peak at the band-touching point due to the effect of two-band quantum geometry.
For consistency checks, we also show the ED results of the $n(\mathbf{k})$ distribution of FCIs in the gapless flat band from different system sizes.
Due to the difficulty of larger-size two-band simulations for the Moore-Read FCI, we take the Laughlin case for example first.
As shown in Fig.~\ref{figs_nk_hc_Laughlin_ED} and Fig.~\ref{fig_nk_adaptiveness}, the boson occupation consistently peaks at the band-touching point.
Moreover, we also show the results from larger-size iDMRG simulations in circumference $N_1=6$ cylinders with hard-core bosons in Fig.~\ref{figs_nk_hc_idmrg}.
For demonstration, we take a region of $6\times6$ unit cells from the infinite cylinders for the Fourier transformation of the correlation functions $\langle b^\dagger_i b^{\ }_j\rangle$, and we have checked that the average momentum-space density matches well with the filling $\bar{n}=\nu$.
Consistently, we observe the peaks of $n(\mathbf{k})$ at the band-touching point, and the peak in the Moore-Read case is more prominent than that in the Laughlin case.

\end{document}